\documentclass[prl,twocolumn,aps,showpacs,floatfix,preprintnumbers,nofootinbib]{revtex4-1}
\usepackage{amsthm}   
\usepackage{amsmath}  
\usepackage{amssymb}  
\usepackage{mathrsfs}
\usepackage{color}
\usepackage[all]{xy}
\usepackage[pdftex]{graphicx}
\usepackage{indentfirst}
\usepackage[pdftex]{hyperref}
\usepackage{subfigure}

\newcommand{\mnras}{Mon.\ Not.\ R.\ Astron.\ Soc.}

\newcommand{\be}{\begin{equation}}
\newcommand{\ee}{\end{equation}}
\newcommand{\ba}{\begin{eqnarray}}
\newcommand{\ea}{\end{eqnarray}}

\newcommand{\dd}[1]{\dot{#1}}

\begin{document}
\title{Limits on Neutrino-Neutrino Scattering in the Early Universe}
\author{Francis-Yan Cyr-Racine}\email{franyan@caltech.edu}
\affiliation{NASA Jet Propulsion Laboratory, California Institute of Technology, Pasadena, CA 91109, USA}
\affiliation{California Institute of Technology, Pasadena, CA 91125, USA}
\author{Kris Sigurdson}\email{krs@phas.ubc.ca}
\affiliation{Department of Physics and Astronomy, University of British Columbia, Vancouver, BC, V6T 1Z1, Canada}

\date{\today}
\begin{abstract}
In the standard model neutrinos are assumed to have streamed across the Universe since they last scattered when the standard-model plasma temperature was $\sim$MeV. The shear stress of free-streaming neutrinos imprints itself gravitationally on the Cosmic Microwave Background (CMB) and makes the CMB a sensitive probe of neutrino scattering. Yet, the presence of nonstandard physics in the neutrino sector may alter this standard chronology and delay neutrino free-streaming until a much later epoch.  We use observations of the CMB to constrain the strength of neutrino self-interactions $G_{\rm eff}$ and put limits on new physics in the neutrino sector from the early Universe. Within the context of conventional $\Lambda$CDM parameters cosmological data are compatible with $G_{\rm eff} \lesssim 1/({\rm 56 \, MeV})^{2}$ and neutrino free-streaming might be delayed until their temperature has cooled to as low as $\sim$25 eV. Intriguingly, we also find an alternative cosmology compatible with cosmological data in which neutrinos scatter off each other until $z\!\sim\!10^4$ with a preferred interaction strength in a narrow region around $G_{\rm eff} \simeq 1/({\rm 10 \, MeV})^{2}\simeq 8.6\times10^8 G_{\rm F}$, where $G_{\rm F}$ is the Fermi constant. This distinct self-interacting neutrino cosmology is characterized by somewhat lower values of both the scalar spectral index and the amplitude of primordial fluctuations. While we phrase our discussion here in terms of a specific scenario, our constraints on the neutrino visibility function are very general. 
\end{abstract}
\maketitle

\section{Introduction}
Neutrinos are the most elusive components of the standard model (SM) of particle physics. Their tremendously weak interactions with other SM fields render measurements of their fundamental properties very challenging. At the same time, the existence of neutrino mass \cite{Ahmad:2001an} constitutes one of the most compelling lines of evidence for physics beyond the SM, and makes the neutrino sector a prime candidate for searches for  new physics. In recent years, cosmology has provided some of the most stringent constraints on neutrino properties, most notably the sum of their masses and their effective number \cite{dePutter:2012sh,*Giusarma:2012ph,planckXVI}. Can cosmological data inform us about other aspects of neutrino physics?

One assumption that is almost always implicitly made 
is the free-streaming nature of cosmological neutrinos (for exceptions, see, e.g. Refs.~\cite{BialynickaBirula:1964zz,Raffelt:1987ah,Chacko:2003dt,Beacom:2004yd, Hannestad:2004qu,Hannestad:2005ex, Bell:2005dr,Sawyer:2006ju,Mangano:2006mp,Friedland:2007vv,Basboll:2008fx,Serra:2009uu,Jeong:2013eza,Archidiacono:2013dua,Archidiacono:2014nda}). Within the confines of the standard model this assumption is justified since SM neutrinos are expected to have decoupled from the primeval plasma in the very early Universe at a temperature $T\simeq{\rm 1.5~MeV}$.   Yet, this assumption is not a priori driven by cosmological observations, but instead a prior on the models of neutrino physics we choose to compare with data. Abandoning this assumption allows us to answer the important question: \emph{How does cosmology inform us about the interactions of neutrinos with each other?}

Free-streaming neutrinos create anisotropic stress which, through gravity, alters the evolution of the other particle species in the Universe \cite{Bashinsky:2003tk,Friedland:2007vv,Hou:2011ec}.
As cosmological fluctuations in the photon and baryon fluids are particularly sensitive to the presence of a free-streaming component during the radiation-dominated era, we expect the recent measurements of the CMB to provide an interesting constraint on the onset of neutrino free-streaming.  We emphasize that while neutrino-neutrino scattering may have been ubiquitous in the early Universe, arranging for and measuring neutrino-neutrino scattering  on Earth is particularly difficult given the challenges involved in creating intense neutrino beams (see e.g.~\cite{Abe:2012av}).

In this paper, we compute the first purely cosmological constraints on the strength of neutrino self-interactions. We model the interaction as a four-fermion vertex whose strength is controlled by a dimensional constant $G_{\nu}$, analogous to the Fermi constant. In this scenario, the onset of neutrino free-streaming is delayed until the rate of these interactions fall below the expansion rate of the Universe, hence affecting the evolution of cosmological fluctuations that enter the causal horizon before that epoch. As we discuss below, the cosmological observables are compatible with a neutrino visibility function peaking at a temperature orders of magnitude below that of the standard picture. Furthermore, we unveil here a novel cosmology in which neutrinos are strongly self-interacting until close to the epoch of matter-radiation equality. 
	
In earlier investigations of neutrino properties \cite{Trotta:2004ty,Melchiorri:2006xs,DeBernardis:2008ys,Smith:2011es,Archidiacono:2011gq,Archidiacono:2012gv,Gerbino:2013ova}, neutrinos were modeled as a fluid-like \cite{Hu:1998kj} and constraints were placed on the phenomenological parameters $c_{\rm eff}$ and $c_{\rm vis}$, the rest-frame sound speed and the viscosity parameter of the neutrino fluid respectively.  These analysis found consistency with the free-streaming limit. However, by modeling these parameters as constant throughout the history of the Universe they could not capture the realistic physics of neutrino decoupling. We incorporate here the physics necessary to follow in detail the dynamics of the transition of neutrinos from a tightly-coupled fluid to particles free-streaming across the Universe. As we discuss below, our analysis shows that the phenomenological parametrization using $c_{\rm eff}$ and $c_{\rm vis}$ has no interpretation in terms of standard particle scattering, hence shedding doubt on the usefulness of these parameters. 

\section{Neutrino Interactions}

%
\subsection{Previous Works and Constraints}
Neutrino interactions beyond those predicted by the SM have been studied in many different contexts over the last three decades. Example of this including Majoron models \cite{Gelmini:1980re,Chikashige:1980qk} in which neutrinos couple to a massless Goldstone boson, as well as scenarios where neutrinos couple to new massive scalar or vector particles \cite{Bardin:1970wq,Barger:1981vd}. In Majoron scenarios where the new boson is a SM gauge singlet, the emission of the massless Goldstone boson in the final state of meson and lepton decays put fairly strong constraints on the dimensionless coupling constant between Majoron and neutrinos (see Ref.~\cite{Lessa:2007up} for recent bounds). We note that scenarios where neutrinos couple to $SU(2)_L$ doublet or triplet Majorons are ruled out by LEP since they would contribute an extra component to the invisible decay width of the $Z$ boson. It was also shown that the CMB places strong constraints on these scenarios \cite{Chacko:2003dt,Beacom:2004yd, Hannestad:2004qu,Hannestad:2005ex,Archidiacono:2013dua,Archidiacono:2014nda}

In models where neutrinos couple to new heavy mediator particles, the low-energy theory relevant to cosmological studies based on the CMB can be described by a dimensionfull Fermi-like constant $G_\nu$. In this type of scenarios, the possible emission of a light (relative to the decaying species) mediator particle by neutrinos in the final state of Kaon, $Z$, and $W$ decay leads to bounds on the value of $G_{\nu}$ \cite{Bardin:1970wq,Barger:1981vd,Bilenky:1992xn,Bilenky:1999dn,Laha:2013xua}. We caution that these bounds must be interpreted with care as they each have their own range of validity and built-in assumptions. For instance, Refs.~\cite{Bilenky:1992xn,Bilenky:1999dn} implicitly assumes the mediator to be at or near the weak scale, making these constraints largely inapplicable for much lighter mediator masses such as those relevant for the present CMB study. On the other hand, Ref.~\cite{Laha:2013xua} considers the case of a light ($\lesssim5$ MeV) vector mediator coupling in a gauge invariant way to both neutrinos and charged leptons, leading to very strong constraints on hidden neutrino interactions. These constraints can however be largely alleviated if the coupling to charged leptons is relaxed, although Big Bang Nucleosynthesis still provides a strong constraint for light mediator mass \cite{Ahlgren:2013wba}.

The copious emission of neutrinos from supernovae provides another avenue to study neutrino interactions beyond the SM \cite{Choi:1989hi,Kachelriess:2000qc,Farzan:2002wx,Davoudiasl:2005fd,Fayet:2006sa,Blennow:2008er,Sher:2011mx,Zhou:2011rc}. It was initially thought that the neutrinos detected from SN1987A place very strong constraints on the neutrino self-interaction cross section \cite{Manohar:1987ec}. However, these bounds were largely refuted in Ref.~\cite{Dicus:1988jh} with the exception of the relatively weak constraints from Ref.~\cite{Kolb:1987qy}. In a similar spirit to this last constraint, the propagation of ultra-high energy neutrinos through cosmological distances also puts bounds on the neutrino self-interaction cross section \cite{Keranen:1997gz,Hooper:2007jr,Ng:2014pca,Ioka:2014kca,Cherry:2014xra}. 

In this work, we present bounds on neutrino interaction that are purely based on the universal gravitational influence of neutrinos on CMB photons. As such, the constraints presented here are largely model independent and thus fully complementary to the limits discussed above. While we use a specific phenomenological scenario to model the neutrino self-interactions (see next section), our constraints on the neutrino visibility are very general. 
\begin{figure}[t]
\begin{center}
\includegraphics[width=0.49\textwidth]{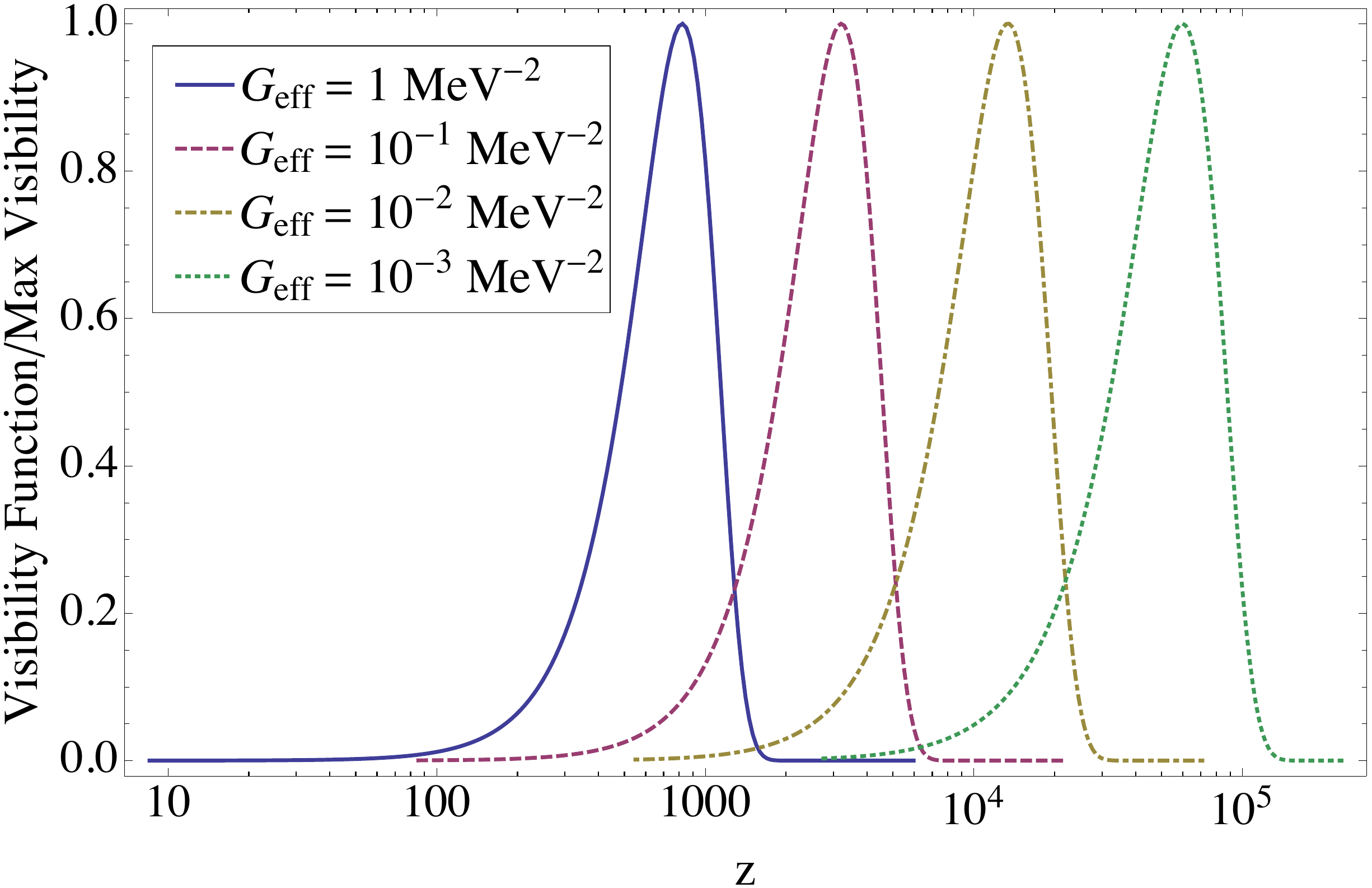}
\caption{Visibility function for self-interacting neutrinos for different values of the effective coupling constant $G_{\rm eff}$ as a function of redshift. Here, we assume three neutrinos species. Note that we have divided out each case by their maximum visibility in order to show them on the same scale.}
\label{visibility}
\end{center}
\end{figure}
\subsection{Effective Scenario}
For definiteness, we consider a phenomenological scenario in which neutrinos, in addition to their regular SM interactions, have non-negligible self-interactions due to their coupling with strength $g_\nu$ to a new massive mediator particle $\varphi$ with mass $M_\varphi$. When the temperature of the neutrinos falls significantly below the mediator mass, one can integrate the latter out and model the interaction as a four-fermion vertex controlled by a dimensionful coupling constant 
\be
G_{\nu}\equiv \frac{g_\nu^2}{M_\varphi^2}.
\ee
As long as neutrinos are relativistic, the thermally-averaged neutrino-neutrino cross section times velocity scales as
\be\label{eq:correspondance}
\langle \sigma_{\nu\nu} v\rangle \propto G_\nu^2T_\nu^2\sim \left(\frac{G_\nu}{G_{\rm F}}\right)^2\langle \sigma_{\nu\nu}^{\rm SM} v\rangle,
\ee
where $T_{\nu}$ is the temperature of the neutrino bath, $G_{\rm F}  \simeq 1.166\times 10^{-11}$ MeV$^{-2}$ is the Fermi constant, $v$ is the velocity of the neutrinos, and where $\sigma_{\nu\nu}^{\rm SM}$ denote the Standard Model neutrino interaction cross section to standard Weak processes. The above relation highlights how quickly the neutrino self-interaction cross section rises for increasing $G_\nu$. In the early Universe, self-interactions render the neutrino medium opaque with an opacity $\dd{\tau}_{\nu}\equiv-a\xi G_{\nu}^2T_{\nu}^5$, where $\xi$ is an order unity constant that depends on the specific neutrino interaction model and on the thermal averaging process, and $a$ is the scale factor describing the expansion of the Universe. We note that $\dd{\tau}_{\nu} \propto \langle \sigma_{\nu\nu} v\rangle$. In this work, we focus our attention on the case where $G_{\nu} \gg G_{\rm F}$. Therefore, we justifiably neglect the contributions from electroweak processes to the neutrino opacity in what follows.  Since the opacity is only sensitive to the product of $\xi$ and $G_\nu^2$, we define an effective coupling constant
\be
G_{\rm eff} \equiv \sqrt{\xi} G_\nu = \sqrt{\xi} \frac{g_\nu^2}{M_\varphi^2}.
\ee
In the following, we shall phrase our constraints uniquely in terms of  $G_{\rm eff}$. The opacity of the neutrino medium implicitly defines a neutrino visibility function given by $\tilde{g}_{\nu}(z)\!\equiv\!-\dd{\tau}_{\nu}e^{-\tau_{\nu}}$. As in the photon case, the visibility function can be thought of as a probability density function for the redshift at which a neutrino begins to free-stream. Compared to the standard case, the introduction of a new type of interaction in the neutrino sector can push the peak of the neutrino visibility function to considerably lower redshift. We illustrate in Fig.~\ref{visibility} the neutrino visibility function for different values of the effective coupling constant $G_{\rm eff}$. 

It is important to keep in mind that in scenarios where neutrinos have extra interactions, we generally expect their thermal history to differ from the standard cosmological scenario (see e.g.~\cite{Boehm:2012gr}). For instance, a new mediator particle could decay or annihilate into neutrinos after they decouple from the SM plasma, hence reheating the neutrino sector compared to the CMB. However, since these types of effects are highly model dependent we do not consider them here and fix the neutrino temperature to the standard value $T_\nu = (4/11)^{1/3}T_{\rm CMB}$ throughout. As we discuss below, for most of the allowed parameter space\footnote{The exception being the peculiar ``interacting neutrino'' cosmology. We refer the reader to the results section for more details.} the mediator particle is heavy enough as to not dramatically change the thermal history of neutrinos. We emphasize that for any realistic model of neutrino self-interaction, it is straightforward to modify our analysis to self-consistently take into account the different thermal history of neutrinos.

\begin{figure*}[t!]
\begin{center}
\includegraphics[width=0.99\textwidth]{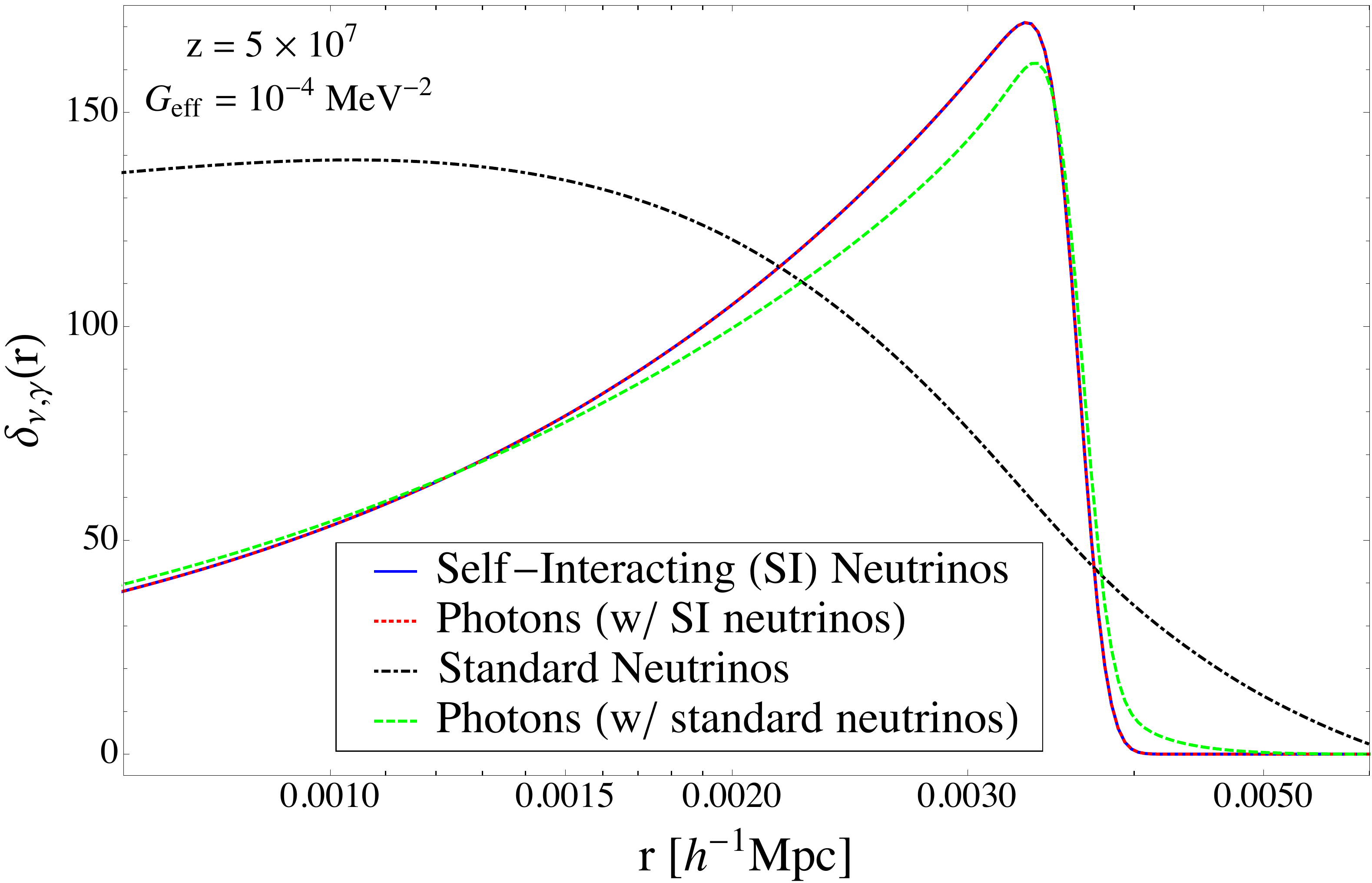}
\vspace{-0.28cm}
\caption{Snapshot of neutrino and photon density fluctuations in configuration space at a fixed redshift. The black dot-dashed line shows the standard free-streaming neutrino fluctuation while the green dashed line displays the corresponding photon density fluctuation.  The solid blue and red dotted lines show the density fluctuation of self-interacting neutrinos and the corresponding photon perturbation, respectively. These two lines lie on top of one another since both neutrinos and photons behave as tightly-coupled fluids at the epoch shown here. The difference between the green dashed and the red dotted lines readily illustrates the phase shift and amplitude suppression of the photon fluctuation associated with free-streaming neutrinos. Here we have adopted a Planck cosmology \cite{planckXVI}.}
\vspace{-0.28cm}
\label{real_space_pert}
\end{center}
\end{figure*}
\begin{figure*}[t]
\begin{center}
\includegraphics[width=0.99\textwidth]{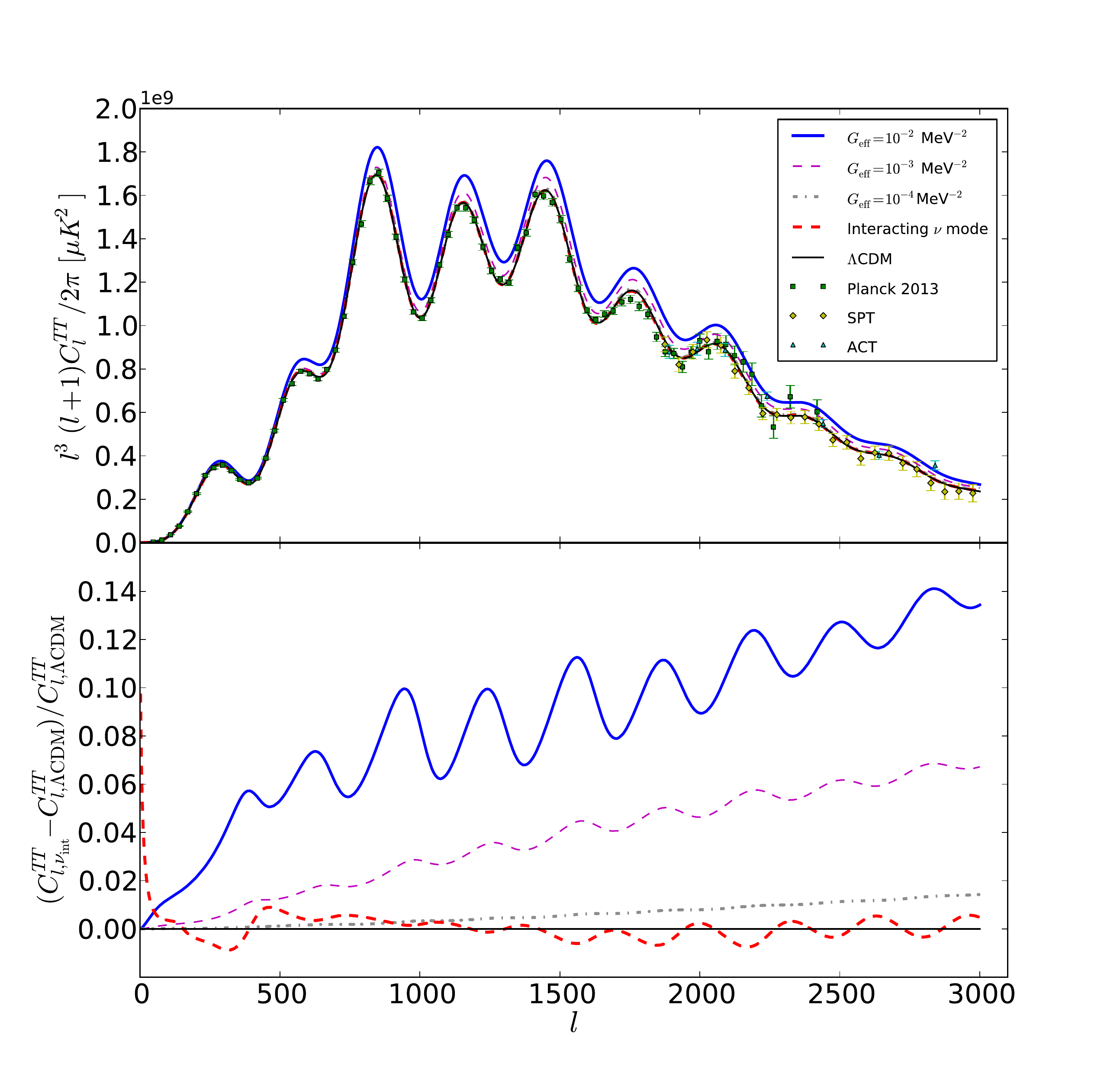}
\vspace{-0.28cm}
\caption{CMB temperature power spectra for different values of $G_{\rm eff}$. The upper panel shows the temperature spectra themselves together with recent measurements, while the lower panel displays the relative difference between the interacting neutrino models and the best-fit Planck $\Lambda$CDM cosmology \cite{planckXVI} with three neutrinos. The dashed red lines illustrate the interacting neutrino cosmology given in Table \ref{table1}. }
\vspace{-0.48cm}
\label{cls}
\end{center}
\end{figure*}
%

\section{Evolution of Cosmological Fluctuations}
To determine the impact of neutrino self-interaction on cosmological observables, we evolve the neutrino fluctuation equations from their early tightly-coupled stage to their late-time free-streaming solution. By prohibiting free-streaming, neutrino self-interactions severely damp the growth of anisotropic stress associated with the quadrupole and higher moments of the neutrino distribution function. Indeed, while the equations for the density and velocity fluctuations of the neutrinos are unaffected by the self-interaction, the moments with $l\geq2$ are corrected by a damping term proportional to $\dd{\tau}_{\nu}$ which effectively suppresses their growth, 
\be\label{Fnu2}
\dd{F}_{\nu2} = \frac{8}{15}\theta_{\nu}+\frac{8}{15}k\sigma-\frac{3}{5}kF_{\nu3}+\alpha_2\dd{\tau}_{\nu}F_{\nu2}
\ee
\be\label{Fnu3}
\dd{F}_{\nu l}=\frac{k}{2l+1}\left[l F_{\nu(l-1)}-(l+1)F_{\nu(l+1)}\right]+\alpha_l\dd{\tau}_{\nu}F_{\nu l},
\ee
where we follow closely the notation of Ref.~\cite{Ma:1995ey} in synchronous gauge. The $\alpha_l$s are order unity $l$-dependent coefficients that depends on the specific model used for neutrino interactions. In our analysis, we set these coefficients to unity; in practice, any change to $\alpha_2$ can be reabsorbed into $G_{\rm eff}$ while changes to $\alpha_l$ for $l\geq3$ have very little impact on the CMB. We note that Eqs.~\ref{Fnu2} and \ref{Fnu3} were derived in the limit that the neutrino distribution function remains thermal throughout decoupling. This approximation is certainly valid at early times when rapid neutrino scattering maintains local thermodynamic equilibrium. As the epoch of neutrino decoupling is approached, the non-negligible momentum transfer in typical neutrino-neutrino collisions can caused the neutrino distribution function to depart from its pure thermal form. We however expect these spectral distortions to be subdominant due to the narrowness of the neutrino visibility function caused by the steep $T_\nu^5$ dependence of the scattering rate. Moreover, the importance of neutrino spectral distortions is further reduced by the fact that neutrinos solely couple to CMB photons via the gravitational potentials, which themselves depend on \emph{integrals} of the neutrino distribution function. While it would be interesting to study and quantify the impact of neutrino spectral distortions on the CMB (see e.g.~\cite{Oldengott:2014qra}), we leave this possibility to future work and assume the form of Eqs.~(\ref{Fnu2}) and (\ref{Fnu3}) to be valid throughout neutrino decoupling.

We solve Eqs.~\ref{Fnu2} and \ref{Fnu3} numerically together with the standard perturbation equations for the photons, baryons and dark matter using a modified version of the code \texttt{CAMB} \cite{Lewis:1999bs}. At early times, the tightly-coupled neutrino equations are very stiff and we use a tight-coupling approximation which sets $F_{\nu2}=8(\theta_{\nu}+k\sigma)/(15\alpha_2\dd{\tau}_{\nu})$ and $F_{\nu l}=0$ for $l\geq3$ \cite{CyrRacine:2010bk}. We note that the neutrino opacity is related to the commonly used viscosity parameter $c_{\rm vis}^2$ though the relation $c_{\rm vis}^2=(1/3)(1-(15/8)\dd{\tau}_{\nu}\alpha_2F_{\nu2}/(\theta_{\nu}+k\sigma))$. As long as neutrinos form a tightly-coupled fluid, the second term is very close to unity and $c_{\rm vis}^2$ approaches zero. After, the onset of neutrino free-streaming, the second term becomes vanishingly small and $c_{\rm vis}^2\rightarrow1/3$. This illustrates that modeling nonstandard neutrino physics with a constant $c_{\rm vis}^2\neq1/3$ has no intuitive meaning in terms of simple particle scattering, hence shedding doubt on the usefulness of this parametrization.

We compare in Fig.~\ref{real_space_pert} the evolution in configuration space of self-interacting and free-streaming neutrino fluctuations. Since it can establish gravitational potential perturbation beyond the sound horizon of the photon-baryon plasma, free-streaming radiation suppresses the amplitude and shift the phase of photon density fluctuations \cite{Bashinsky:2003tk,Friedland:2007vv,Hou:2011ec}. For each Fourier mode of the photon fluctuations, the magnitude of these two effects is directly proportional to the free-streaming fraction of the total radiation energy density when the Fourier mode enters the Hubble horizon. If neutrino free-streaming is delayed due to their self-interaction until redshift $z_{\nu*}$, Fourier modes of photon fluctuations entering the horizon before $z_{\nu*}$ would not be affected by the standard shift in amplitude and phase. On the other hand, the amplitude of photon fluctuations becoming sub-horizon at a redshift  $z_{\rm eq}\!<\!z\!<\!z_{\nu*}$ would be suppressed and their phase would be shifted toward larger scales (smaller $l$). Therefore, the impact of delayed neutrino free-streaming on the temperature and polarization power spectra of the CMB is a $l$-dependent shift in their amplitude and phase. Multipoles with $l_{\rm eq}\!<\!l\!<\!l_{\nu*}$ are largely unaffected by neutrino self-interaction while multipoles with $l\!>\!l_{\nu*}$ are expected to gradually display more power and have their phase shifted toward smaller angular scales as $l$ is increased. We illustrate these signatures of neutrino self-interaction for different values of $G_{\rm eff}$ in Fig.~\ref{cls}.

\section{Data}
To constrain neutrino self-interaction, we use the CMB data from the Planck satellite \cite{planckXVI}. We utilize both the low-multipole and high-multipole temperature data, incorporating the required ``nuisance'' parameters describing foregrounds and instrumental effects, and also include the WMAP low-$l$ polarization data.  We refer to this dataset as ``Planck+WP''. We also incorporate the high-resolution temperature data from the South Pole Telescope (SPT) and the Atacama Cosmology Telescope (ACT). As in the original Planck analysis, we only include the ACT $148\!\times\!148$ spectra for $l\!\geq\!1000$, the ACT $148\!\times\!218$ and $218\!\times\!218$ spectra for $l\!\geq\!1500$ \cite{Dunkley:2013vu,Das:2013zf}, and the SPT data described in \cite{Story:2012wx} for $l\!\geq\!2000$. We fully incorporate the nuisance parameters describing foregrounds and calibration uncertainties for both SPT and ACT. We collectively refer to this dataset as ``High-$l$''. 

We also include in our analysis baryon acoustic oscillation (BAO) data from a reanalysis of the Sloan Digital Sky Survey DR7 \cite{Padmanabhan:2012hf}, from the 6-degree Field survey \cite{Beutler:2011hx}, and from the Baryon Oscillation Spectroscopic Survey \cite{Anderson:2012sa}. For our cosmological parameter estimation, we use the publicly available Markov Chain Monte Carlo code \texttt{CosmoMC} \cite{Lewis:2002ah}.
We also obtain a pre-Planck era constraint on neutrino self-interaction by using WMAP9 temperature and polarization data \cite{Hinshaw:2012aka} in addition to the high-resolution temperature data from SPT and ACT. For this analysis, we use the ACT temperature data from the equatorial patch for $500\!<\!l\!<\!3500$ and SPT temperature data for $650\!<\!l\!<\!3000$ as described in \cite{Calabrese:2013jyk}. In both cases, these spectra are pre-calibrated to WMAP and pre-marginalized over foregrounds. We collectively refer to this dataset as ``WMAP9 + ACT + SPT''. While the cosmological results from this last combination of datasets are somewhat in tension with those determined by Planck, we will see that our results are robust and only weakly depend on the specific datasets considered. 

\section{Results and Discussion}
We run Markov Chain Monte Carlo analyses with the above-mentioned data, letting the standard six parameters of $\Lambda$CDM vary ($\Omega_{\rm c}h^2$, $\Omega_{\rm b}h^2$,$\theta_*$,$\tau$,$n_{\rm s}$ and $\ln{(10^{10}A_{\rm s})}$) in addition to varying $G_{\rm eff}$ and the nuisance parameters. We set the prior distributions to those described in \cite{planckXVI}, and use a flat prior on $\log_{10}(G_{\rm eff} {\rm MeV}^2)\!\in\!\left[-6,0\right]$ (we will relax this assumption below). To ensure that we fully explore the posterior distribution, we generate Markov chains at high temperature and obtain our final posterior by importance sampling. To verify chain convergence, we run ten independent chains for each combination of data sets and make sure that each yields a similar posterior, and that the Gelman-Rubin convergence criterion between the chains is at most $R-1=0.02$ for the least converged parameter. In our analysis, we fix the effective number of neutrinos to the standard value of 3.046 and focus on massless neutrinos. We will expand our analysis to massive neutrinos in future work. 

\begin{figure}[h!]
\begin{center}
\includegraphics[width=0.49\textwidth]{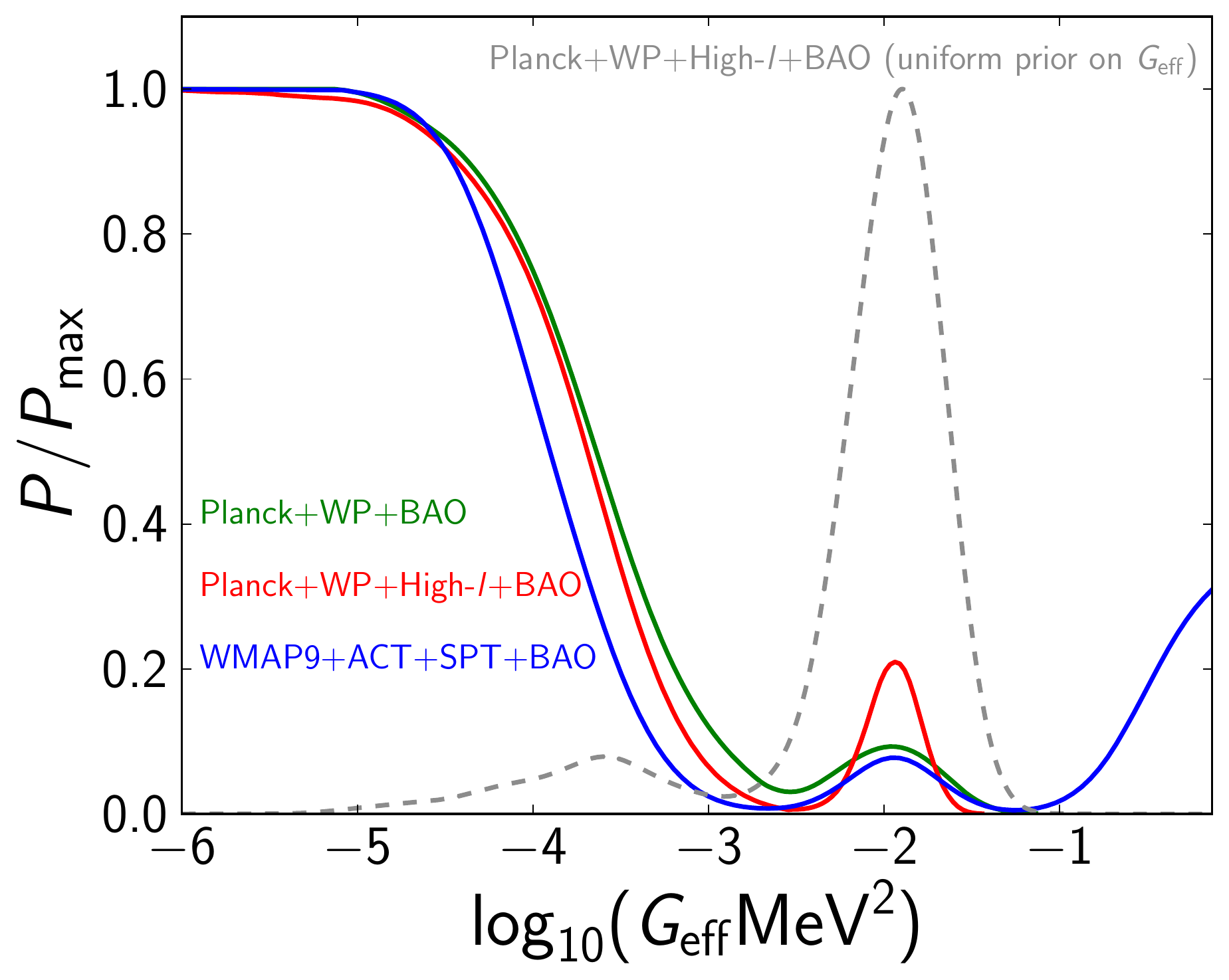}
\includegraphics[width=0.49\textwidth]{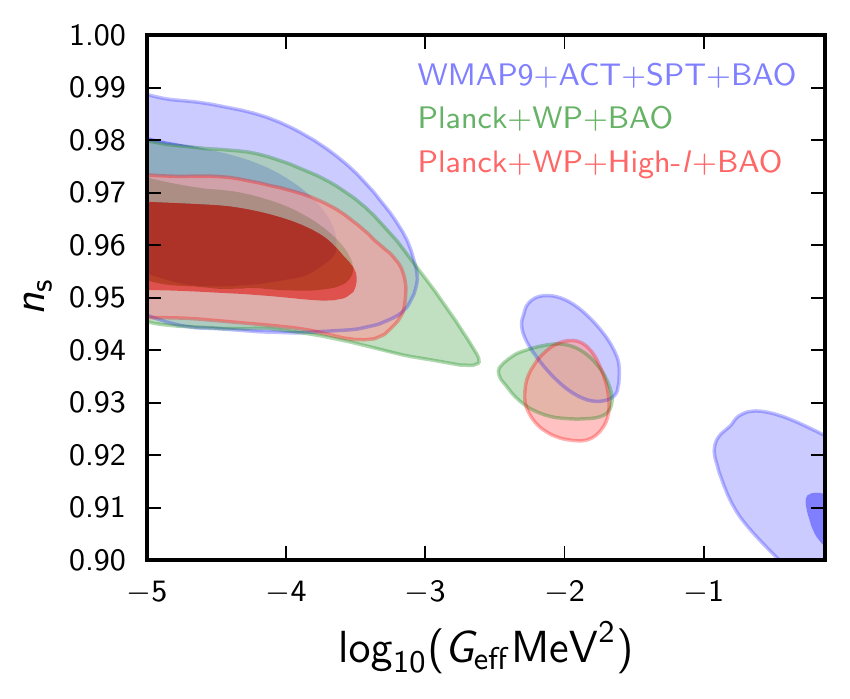}
\caption{{\it Top panel:} Marginalized posterior distribution of $\log_{10}({\rm G_{\rm eff}MeV}^2)$ for different combinations of datasets and priors. {\it Bottom panel:} 2D marginalized constraints in the $n_{\rm s}$  and $\log_{10}({\rm G_{\rm eff}MeV}^2)$ plane.}
\label{logGeff_1D}
\end{center}
\end{figure}

We show in Fig.~\ref{logGeff_1D} the marginalized posterior distribution of $\log_{10}(G_{\rm eff}{\rm MeV}^2)$ for all the combinations of datasets considered. We surprisingly observe that the marginalized posterior is \emph{multimodal}. To avoid quoting misleading bounds, we provide below confidence intervals for each mode separately. It is important to emphasize that the posterior distribution of nuisance parameters is not affected by the introduction of $G_{\rm eff}$, indicating that the effect of neutrino interaction is not degenerate with foreground contamination and calibration uncertainties.  
\subsection{Standard Cosmological Mode}
The principal mode of the distribution, which connects continuously with the standard cosmological scenario, spans the range $G_{\rm eff} \leq 10^{-2.6}{\rm MeV}^{-2}$. For this mode, the exact confidence intervals strongly depend on the lower limit of the flat prior on $\log_{10}(G_{\rm eff}{\rm MeV}^2)$ since $G_{\rm eff} \lesssim10^{-5}\,{\rm MeV}^{-2}$ has little impact on the CMB. For our choice of prior, we obtain $\log_{10}(G_{\rm eff}{\rm MeV}^2)\! <\! -3.5$ ($95\%$ C.L.) for ``Planck+WP+High-$l$+BAO''. In terms of the Fermi constant, this bound corresponds to  $\log_{10}(G_{\rm eff}/G_{\rm F})\! <\! 7.4$ ($95\%$ C.L.). Assuming a scalar mediator which saturates the coupling strength limit of Refs.~\cite{Barger:1981vd,Lessa:2007up}, this bound implies a mediator mass $M_\varphi\gtrsim0.4$ MeV.  Within this mode, the range of allowed $G_{\rm eff}$ values is remarkably large, implying that the onset of neutrino free-streaming could have been significantly delayed beyond weak decoupling without affecting cosmological observables. Recasting the above limit into a model-independent lower bound on the peak of the neutrino visibility function, we obtain $z_{\nu*} \gtrsim 1.5 \times 10^5$. This in turns implies that the temperature of the cosmological neutrino bath at the onset of free-streaming could have been as low as $\sim25$ eV. It is important to emphasize that this number is almost \emph{5 orders of magnitude} below the standard value of $T_{\nu,{\rm dec}}\simeq1.5$~MeV. While this observation \emph{does not} imply the presence of new physics in the neutrino sector, it \emph{does} show that there is considerable room for new physics to turn up in the way neutrinos interact.
\subsection{Interacting Neutrino Mode}
The secondary mode of the posterior distribution, which spans $10^{-2.6}\!\!<\!\!G_{\rm eff}{\rm MeV}^{2}\!\!<\!\!10^{-1.3}$, represents a truly novel cosmological scenario. In this ``interacting neutrino'' cosmology, neutrinos are tightly-coupled until $z_{\nu*}\!\sim\!10^4$ such that most of the CMB multipoles do not receive the usual phase shift and amplitude suppression associated with the presence of free-streaming radiation. The presence of this new mode with $\log_{10}(G_{\rm eff} {\rm MeV}^2)=-2.0\pm0.2$ ($68\%$ C.L.) or equivalently, $\log_{10}(G_{\rm eff} /G_{\rm F})=8.9\pm0.2$, indicates that the absence of these ``free-streaming'' effects can be compensated by adjusting the other cosmological parameters, especially the scalar spectral index and the amplitude of primordial fluctuations (see Table \ref{table1}). This points to a previously unknown degeneracy between the spectrum of primordial fluctuations and the gravitational effect of the neutrinos on the CMB. We display in Fig.~\ref{cls} (red dashed line) an example of the CMB temperature spectrum for the interacting neutrino cosmology. We observe that the interacting neutrino cosmology temperature spectrum closely follows the best-fit $\Lambda$CDM Planck spectrum, except at the lowest multipoles where sample variance dominates. We note that the error bars of the WMAP 9-year data allows for an additional mode of non-vanishing probability at large $G_{\rm eff}$ values. This region is  disfavored by current Planck data and we do not further consider this region of parameter space. 
\begin{table}[t]
\begin{center}
\begin{tabular}{|c|c|c|}
\hline
Parameters & Standard Mode & Interacting-$\nu$ Mode \\
\hline
\hline
$\Omega_{\rm b}h^2$ & $0.0221\pm0.0002$ & $0.0222\pm0.0003$\\
$\Omega_{\rm c}h^2$  & $0.119\pm0.002$ & $0.120\pm0.002$  \\
$\tau$  & $0.09\pm0.01$ &  $0.09\pm0.01$\\
$H_0$ & $68.1\pm0.8$ & $69.0\pm0.8$ \\
$n_{\rm s}$ &$0.959\pm0.006$&$0.932\pm0.006$ \\
$10^9A_{\rm s}$ & $2.19\pm0.02$ & $2.07\pm0.02$ \\
$\log_{10}(G_{\rm eff}{\rm MeV}^2)$ & $< -3.5\,(95 \%\,{\rm C.L.})$ & $-2.0\pm0.2$\\
$\log_{10}(G_{\rm eff}/G_{\rm F})$ & $< 7.4\,(95 \%\,{\rm C.L.})$ & $8.9\pm0.2$\\
\hline
\end{tabular}
\caption{Marginalized constraints on cosmological parameters for the two main modes of the distribution. Unless otherwise indicated, we quote $68\%$ confidence level for Planck + WP + High-$l$ + BAO.\label{table1}}
\end{center}
\end{table}

How significant is the interacting neutrino cosmology? From Fig.~\ref{logGeff_1D}, it is clear that the weight of the interacting neutrino mode in the posterior is smaller compared to that of the principal mode. This is however the result of our choice of prior: a uniform prior on $\log_{10}(G_{\rm eff} {\rm MeV}^2)$ is equivalent to setting a non-uniform prior on $G_{\rm eff}$ which scales as $1/G_{\rm eff}$. Our choice of prior thus gives larger weights to small values of $G_{\rm eff}$, hence favoring the standard $\Lambda$CDM model. However, if we instead impose (an arguably equally reasonable) uniform prior on $G_{\rm eff}$, then the interacting neutrino cosmology becomes favored over the standard cosmological model (see dashed line in top panel of Fig.~\ref{logGeff_1D}). Therefore, it is clear that additional datasets will have to be considered to determine whether the interacting neutrino cosmology is a plausible scenario. Of course, it is important to keep in mind that such a large effective coupling constant corresponds to a neutrino interaction cross section that is more than 17 orders of magnitude above that of the SM (see Eq.~(\ref{eq:correspondance}) above). Within our simple phenomenological framework, such a large cross section requires the mass of the mediator to be very light ($\sim 50$ keV) which is likely in conflict with BBN bounds on the number of relativistic species in the early Universe \cite{Ahlgren:2013wba}. However, it remains to be seen whether more complex (and realistic) models of neutrino interactions (see, e.g.,~\cite{Hannestad:2013ana,Dasgupta:2013zpn,Bringmann:2013vra,Cherry:2014xra}) can be built to accommodate such a large interaction cross section while evading other constraints.  It is nevertheless intriguing that this alternate cosmology is only viable for a narrow range of the neutrino interaction strength. 
\section{Conclusion}
In conclusion, we have shown that the CMB allows for a neutrino self-interaction strength that is orders of magnitude larger than the standard Fermi constant. Moreover, we have determined that strongly self-interacting neutrinos with $G_{\rm eff} \simeq 1/({\rm 10 \, MeV})^{2}\simeq8.6\times10^8 G_{\rm F} $ can lead to a CMB spectrum that is in very good agreement with the data. We expect that upcoming CMB polarization data from Planck, SPTpol \cite{Austermann:2012ga}, and ACTpol \cite{Niemack:2010wz} will slightly improve our limits by providing independent constraints on the phase shift and amplitude change caused by self-interacting neutrinos. Moreover, CMB lensing reconstruction will likely provide tight constraints on the interacting neutrino cosmology by narrowing the allowed range of $n_{\rm s}$ and $A_{\rm s}$. Given the relatively large interaction strengths discussed here, it is interesting to consider whether tests of self-interacting neutrino physics might be made with extensions of existing neutrino beam experiments (see, e.g., Refs.~\cite{Hylen:2002uc,Abe:2012av}), a rather exciting possibility.  

{\noindent \bf Acknowledgements---} We thank Roland de Putter and Olivier Dor\'e for useful discussions. The work of FYCR is supported by the W.M. Keck Institute for Space Studies. The research of KS is supported in part by a National Science and Engineering Research Council (NSERC) of Canada Discovery Grant.  We thank the Kavli Institute for Theoretical Physics, where part of this work was completed, for their hospitality.  This research was supported in part by the National Science Foundation under Grant No. NSF PHY11-25915.  Part of the research described in this paper was carried out at the Jet Propulsion Laboratory, California Institute of Technology, under a contract with the National Aeronautics and Space Administration.

\bibliography{Interacting_neutrinos}

\begin{thebibliography}{75}%
\makeatletter
\providecommand \@ifxundefined [1]{%
 \@ifx{#1\undefined}
}%
\providecommand \@ifnum [1]{%
 \ifnum #1\expandafter \@firstoftwo
 \else \expandafter \@secondoftwo
 \fi
}%
\providecommand \@ifx [1]{%
 \ifx #1\expandafter \@firstoftwo
 \else \expandafter \@secondoftwo
 \fi
}%
\providecommand \natexlab [1]{#1}%
\providecommand \enquote  [1]{``#1''}%
\providecommand \bibnamefont  [1]{#1}%
\providecommand \bibfnamefont [1]{#1}%
\providecommand \citenamefont [1]{#1}%
\providecommand \href@noop [0]{\@secondoftwo}%
\providecommand \href [0]{\begingroup \@sanitize@url \@href}%
\providecommand \@href[1]{\@@startlink{#1}\@@href}%
\providecommand \@@href[1]{\endgroup#1\@@endlink}%
\providecommand \@sanitize@url [0]{\catcode `\\12\catcode `\$12\catcode
  `\&12\catcode `\#12\catcode `\^12\catcode `\_12\catcode `\%12\relax}%
\providecommand \@@startlink[1]{}%
\providecommand \@@endlink[0]{}%
\providecommand \url  [0]{\begingroup\@sanitize@url \@url }%
\providecommand \@url [1]{\endgroup\@href {#1}{\urlprefix }}%
\providecommand \urlprefix  [0]{URL }%
\providecommand \Eprint [0]{\href }%
\providecommand \doibase [0]{http://dx.doi.org/}%
\providecommand \selectlanguage [0]{\@gobble}%
\providecommand \bibinfo  [0]{\@secondoftwo}%
\providecommand \bibfield  [0]{\@secondoftwo}%
\providecommand \translation [1]{[#1]}%
\providecommand \BibitemOpen [0]{}%
\providecommand \bibitemStop [0]{}%
\providecommand \bibitemNoStop [0]{.\EOS\space}%
\providecommand \EOS [0]{\spacefactor3000\relax}%
\providecommand \BibitemShut  [1]{\csname bibitem#1\endcsname}%
\let\auto@bib@innerbib\@empty
\bibitem [{\citenamefont {Ahmad}\ \emph {et~al.}(2001)\citenamefont {Ahmad}
  \emph {et~al.}}]{Ahmad:2001an}%
  \BibitemOpen
  \bibfield  {author} {\bibinfo {author} {\bibfnamefont {Q.}~\bibnamefont
  {Ahmad}} \emph {et~al.} (\bibinfo {collaboration} {SNO Collaboration}),\
  }\href {\doibase 10.1103/PhysRevLett.87.071301} {\bibfield  {journal}
  {\bibinfo  {journal} {Phys.Rev.Lett.}\ }\textbf {\bibinfo {volume} {87}},\
  \bibinfo {pages} {071301} (\bibinfo {year} {2001})}\BibitemShut {NoStop}%
\bibitem [{\citenamefont {de~Putter}\ \emph {et~al.}(2012)\citenamefont
  {de~Putter}, \citenamefont {Mena}, \citenamefont {Giusarma}, \citenamefont
  {Ho} \emph {et~al.}}]{dePutter:2012sh}%
  \BibitemOpen
  \bibfield  {author} {\bibinfo {author} {\bibfnamefont {R.}~\bibnamefont
  {de~Putter}}, \bibinfo {author} {\bibfnamefont {O.}~\bibnamefont {Mena}},
  \bibinfo {author} {\bibfnamefont {E.}~\bibnamefont {Giusarma}}, \bibinfo
  {author} {\bibfnamefont {S.}~\bibnamefont {Ho}},  \emph {et~al.},\ }\href
  {\doibase 10.1088/0004-637X/761/1/12} {\bibfield  {journal} {\bibinfo
  {journal} {Astrophys.J.}\ }\textbf {\bibinfo {volume} {761}},\ \bibinfo
  {pages} {12} (\bibinfo {year} {2012})}\BibitemShut {NoStop}%
\bibitem [{\citenamefont {Giusarma}\ \emph {et~al.}(2013)\citenamefont
  {Giusarma}, \citenamefont {de~Putter},\ and\ \citenamefont
  {Mena}}]{Giusarma:2012ph}%
  \BibitemOpen
  \bibfield  {author} {\bibinfo {author} {\bibfnamefont {E.}~\bibnamefont
  {Giusarma}}, \bibinfo {author} {\bibfnamefont {R.}~\bibnamefont {de~Putter}},
  \ and\ \bibinfo {author} {\bibfnamefont {O.}~\bibnamefont {Mena}},\ }\href
  {\doibase 10.1103/PhysRevD.87.043515} {\bibfield  {journal} {\bibinfo
  {journal} {Phys.Rev.}\ }\textbf {\bibinfo {volume} {D87}},\ \bibinfo {pages}
  {043515} (\bibinfo {year} {2013})}\BibitemShut {NoStop}%
\bibitem [{\citenamefont {Ade}\ \emph {et~al.}(2013)\citenamefont {Ade} \emph
  {et~al.}}]{planckXVI}%
  \BibitemOpen
  \bibfield  {author} {\bibinfo {author} {\bibfnamefont {P.}~\bibnamefont
  {Ade}} \emph {et~al.} (\bibinfo {collaboration} {Planck Collaboration}),\
  }\href@noop {} {\  (\bibinfo {year} {2013})},\ \Eprint
  {http://arxiv.org/abs/1303.5076} {arXiv:1303.5076 [astro-ph.CO]} \BibitemShut
  {NoStop}%
\bibitem [{\citenamefont {Bialynicka-Birula}(1964)}]{BialynickaBirula:1964zz}%
  \BibitemOpen
  \bibfield  {author} {\bibinfo {author} {\bibfnamefont {Z.}~\bibnamefont
  {Bialynicka-Birula}},\ }\href {\doibase 10.1007/BF02749481} {\bibfield
  {journal} {\bibinfo  {journal} {Nuovo Cim.}\ }\textbf {\bibinfo {volume}
  {33}},\ \bibinfo {pages} {1484} (\bibinfo {year} {1964})}\BibitemShut
  {NoStop}%
\bibitem [{\citenamefont {Raffelt}\ and\ \citenamefont
  {Silk}(1987)}]{Raffelt:1987ah}%
  \BibitemOpen
  \bibfield  {author} {\bibinfo {author} {\bibfnamefont {G.}~\bibnamefont
  {Raffelt}}\ and\ \bibinfo {author} {\bibfnamefont {J.}~\bibnamefont {Silk}},\
  }\href {\doibase 10.1016/0370-2693(87)91143-9} {\bibfield  {journal}
  {\bibinfo  {journal} {Phys.Lett.}\ }\textbf {\bibinfo {volume} {B192}},\
  \bibinfo {pages} {65} (\bibinfo {year} {1987})}\BibitemShut {NoStop}%
\bibitem [{\citenamefont {Chacko}\ \emph {et~al.}(2004)\citenamefont {Chacko},
  \citenamefont {Hall}, \citenamefont {Okui},\ and\ \citenamefont
  {Oliver}}]{Chacko:2003dt}%
  \BibitemOpen
  \bibfield  {author} {\bibinfo {author} {\bibfnamefont {Z.}~\bibnamefont
  {Chacko}}, \bibinfo {author} {\bibfnamefont {L.~J.}\ \bibnamefont {Hall}},
  \bibinfo {author} {\bibfnamefont {T.}~\bibnamefont {Okui}}, \ and\ \bibinfo
  {author} {\bibfnamefont {S.~J.}\ \bibnamefont {Oliver}},\ }\href {\doibase
  10.1103/PhysRevD.70.085008} {\bibfield  {journal} {\bibinfo  {journal}
  {Phys.Rev.}\ }\textbf {\bibinfo {volume} {D70}},\ \bibinfo {pages} {085008}
  (\bibinfo {year} {2004})}\BibitemShut {NoStop}%
\bibitem [{\citenamefont {Beacom}\ \emph {et~al.}(2004)\citenamefont {Beacom},
  \citenamefont {Bell},\ and\ \citenamefont {Dodelson}}]{Beacom:2004yd}%
  \BibitemOpen
  \bibfield  {author} {\bibinfo {author} {\bibfnamefont {J.~F.}\ \bibnamefont
  {Beacom}}, \bibinfo {author} {\bibfnamefont {N.~F.}\ \bibnamefont {Bell}}, \
  and\ \bibinfo {author} {\bibfnamefont {S.}~\bibnamefont {Dodelson}},\ }\href
  {\doibase 10.1103/PhysRevLett.93.121302} {\bibfield  {journal} {\bibinfo
  {journal} {Phys.Rev.Lett.}\ }\textbf {\bibinfo {volume} {93}},\ \bibinfo
  {pages} {121302} (\bibinfo {year} {2004})}\BibitemShut {NoStop}%
\bibitem [{\citenamefont {Hannestad}(2005)}]{Hannestad:2004qu}%
  \BibitemOpen
  \bibfield  {author} {\bibinfo {author} {\bibfnamefont {S.}~\bibnamefont
  {Hannestad}},\ }\href {\doibase 10.1088/1475-7516/2005/02/011} {\bibfield
  {journal} {\bibinfo  {journal} {JCAP}\ }\textbf {\bibinfo {volume} {0502}},\
  \bibinfo {pages} {011} (\bibinfo {year} {2005})}\BibitemShut {NoStop}%
\bibitem [{\citenamefont {Hannestad}\ and\ \citenamefont
  {Raffelt}(2005)}]{Hannestad:2005ex}%
  \BibitemOpen
  \bibfield  {author} {\bibinfo {author} {\bibfnamefont {S.}~\bibnamefont
  {Hannestad}}\ and\ \bibinfo {author} {\bibfnamefont {G.}~\bibnamefont
  {Raffelt}},\ }\href {\doibase 10.1103/PhysRevD.72.103514} {\bibfield
  {journal} {\bibinfo  {journal} {Phys.Rev.}\ }\textbf {\bibinfo {volume}
  {D72}},\ \bibinfo {pages} {103514} (\bibinfo {year} {2005})},\ \Eprint
  {http://arxiv.org/abs/hep-ph/0509278} {arXiv:hep-ph/0509278 [hep-ph]}
  \BibitemShut {NoStop}%
\bibitem [{\citenamefont {Bell}\ \emph {et~al.}(2006)\citenamefont {Bell},
  \citenamefont {Pierpaoli},\ and\ \citenamefont {Sigurdson}}]{Bell:2005dr}%
  \BibitemOpen
  \bibfield  {author} {\bibinfo {author} {\bibfnamefont {N.~F.}\ \bibnamefont
  {Bell}}, \bibinfo {author} {\bibfnamefont {E.}~\bibnamefont {Pierpaoli}}, \
  and\ \bibinfo {author} {\bibfnamefont {K.}~\bibnamefont {Sigurdson}},\ }\href
  {\doibase 10.1103/PhysRevD.73.063523} {\bibfield  {journal} {\bibinfo
  {journal} {Phys.Rev.}\ }\textbf {\bibinfo {volume} {D73}},\ \bibinfo {pages}
  {063523} (\bibinfo {year} {2006})}\BibitemShut {NoStop}%
\bibitem [{\citenamefont {Sawyer}(2006)}]{Sawyer:2006ju}%
  \BibitemOpen
  \bibfield  {author} {\bibinfo {author} {\bibfnamefont {R.~F.}\ \bibnamefont
  {Sawyer}},\ }\href {\doibase 10.1103/PhysRevD.74.043527} {\bibfield
  {journal} {\bibinfo  {journal} {Phys.Rev.}\ }\textbf {\bibinfo {volume}
  {D74}},\ \bibinfo {pages} {043527} (\bibinfo {year} {2006})}\BibitemShut
  {NoStop}%
\bibitem [{\citenamefont {Mangano}\ \emph {et~al.}(2006)\citenamefont
  {Mangano}, \citenamefont {Melchiorri}, \citenamefont {Serra}, \citenamefont
  {Cooray},\ and\ \citenamefont {Kamionkowski}}]{Mangano:2006mp}%
  \BibitemOpen
  \bibfield  {author} {\bibinfo {author} {\bibfnamefont {G.}~\bibnamefont
  {Mangano}}, \bibinfo {author} {\bibfnamefont {A.}~\bibnamefont {Melchiorri}},
  \bibinfo {author} {\bibfnamefont {P.}~\bibnamefont {Serra}}, \bibinfo
  {author} {\bibfnamefont {A.}~\bibnamefont {Cooray}}, \ and\ \bibinfo {author}
  {\bibfnamefont {M.}~\bibnamefont {Kamionkowski}},\ }\href {\doibase
  10.1103/PhysRevD.74.043517} {\bibfield  {journal} {\bibinfo  {journal}
  {Phys.Rev.}\ }\textbf {\bibinfo {volume} {D74}},\ \bibinfo {pages} {043517}
  (\bibinfo {year} {2006})}\BibitemShut {NoStop}%
\bibitem [{\citenamefont {Friedland}\ \emph {et~al.}(2007)\citenamefont
  {Friedland}, \citenamefont {Zurek},\ and\ \citenamefont
  {Bashinsky}}]{Friedland:2007vv}%
  \BibitemOpen
  \bibfield  {author} {\bibinfo {author} {\bibfnamefont {A.}~\bibnamefont
  {Friedland}}, \bibinfo {author} {\bibfnamefont {K.~M.}\ \bibnamefont
  {Zurek}}, \ and\ \bibinfo {author} {\bibfnamefont {S.}~\bibnamefont
  {Bashinsky}},\ }\href@noop {} {\  (\bibinfo {year} {2007})},\ \Eprint
  {http://arxiv.org/abs/0704.3271} {arXiv:0704.3271 [astro-ph]} \BibitemShut
  {NoStop}%
\bibitem [{\citenamefont {Basboll}\ \emph {et~al.}(2009)\citenamefont
  {Basboll}, \citenamefont {Bjaelde}, \citenamefont {Hannestad},\ and\
  \citenamefont {Raffelt}}]{Basboll:2008fx}%
  \BibitemOpen
  \bibfield  {author} {\bibinfo {author} {\bibfnamefont {A.}~\bibnamefont
  {Basboll}}, \bibinfo {author} {\bibfnamefont {O.~E.}\ \bibnamefont
  {Bjaelde}}, \bibinfo {author} {\bibfnamefont {S.}~\bibnamefont {Hannestad}},
  \ and\ \bibinfo {author} {\bibfnamefont {G.~G.}\ \bibnamefont {Raffelt}},\
  }\href {\doibase 10.1103/PhysRevD.79.043512} {\bibfield  {journal} {\bibinfo
  {journal} {Phys.Rev.}\ }\textbf {\bibinfo {volume} {D79}},\ \bibinfo {pages}
  {043512} (\bibinfo {year} {2009})}\BibitemShut {NoStop}%
\bibitem [{\citenamefont {Serra}\ \emph {et~al.}(2010)\citenamefont {Serra},
  \citenamefont {Zalamea}, \citenamefont {Cooray}, \citenamefont {Mangano},\
  and\ \citenamefont {Melchiorri}}]{Serra:2009uu}%
  \BibitemOpen
  \bibfield  {author} {\bibinfo {author} {\bibfnamefont {P.}~\bibnamefont
  {Serra}}, \bibinfo {author} {\bibfnamefont {F.}~\bibnamefont {Zalamea}},
  \bibinfo {author} {\bibfnamefont {A.}~\bibnamefont {Cooray}}, \bibinfo
  {author} {\bibfnamefont {G.}~\bibnamefont {Mangano}}, \ and\ \bibinfo
  {author} {\bibfnamefont {A.}~\bibnamefont {Melchiorri}},\ }\href {\doibase
  10.1103/PhysRevD.81.043507} {\bibfield  {journal} {\bibinfo  {journal}
  {Phys.Rev.}\ }\textbf {\bibinfo {volume} {D81}},\ \bibinfo {pages} {043507}
  (\bibinfo {year} {2010})}\BibitemShut {NoStop}%
\bibitem [{\citenamefont {Jeong}\ and\ \citenamefont
  {Takahashi}(2013)}]{Jeong:2013eza}%
  \BibitemOpen
  \bibfield  {author} {\bibinfo {author} {\bibfnamefont {K.~S.}\ \bibnamefont
  {Jeong}}\ and\ \bibinfo {author} {\bibfnamefont {F.}~\bibnamefont
  {Takahashi}},\ }\href {\doibase 10.1016/j.physletb.2013.07.001} {\bibfield
  {journal} {\bibinfo  {journal} {Phys. \ Lett. \ B {\bf 725},}\ }\textbf
  {\bibinfo {volume} {134}} (\bibinfo {year} {2013}),\
  10.1016/j.physletb.2013.07.001},\ \Eprint {http://arxiv.org/abs/1305.6521}
  {arXiv:1305.6521 [hep-ph]} \BibitemShut {NoStop}%
\bibitem [{\citenamefont {Archidiacono}\ and\ \citenamefont
  {Hannestad}(2014)}]{Archidiacono:2013dua}%
  \BibitemOpen
  \bibfield  {author} {\bibinfo {author} {\bibfnamefont {M.}~\bibnamefont
  {Archidiacono}}\ and\ \bibinfo {author} {\bibfnamefont {S.}~\bibnamefont
  {Hannestad}},\ }\href {\doibase 10.1088/1475-7516/2014/07/046} {\bibfield
  {journal} {\bibinfo  {journal} {JCAP}\ }\textbf {\bibinfo {volume} {1407}},\
  \bibinfo {pages} {046} (\bibinfo {year} {2014})},\ \Eprint
  {http://arxiv.org/abs/1311.3873} {arXiv:1311.3873 [astro-ph.CO]} \BibitemShut
  {NoStop}%
\bibitem [{\citenamefont {Archidiacono}\ \emph {et~al.}(2014)\citenamefont
  {Archidiacono}, \citenamefont {Hannestad}, \citenamefont {Hansen},\ and\
  \citenamefont {Tram}}]{Archidiacono:2014nda}%
  \BibitemOpen
  \bibfield  {author} {\bibinfo {author} {\bibfnamefont {M.}~\bibnamefont
  {Archidiacono}}, \bibinfo {author} {\bibfnamefont {S.}~\bibnamefont
  {Hannestad}}, \bibinfo {author} {\bibfnamefont {R.~S.}\ \bibnamefont
  {Hansen}}, \ and\ \bibinfo {author} {\bibfnamefont {T.}~\bibnamefont
  {Tram}},\ }\href@noop {} {\  (\bibinfo {year} {2014})},\ \Eprint
  {http://arxiv.org/abs/1404.5915} {arXiv:1404.5915 [astro-ph.CO]} \BibitemShut
  {NoStop}%
\bibitem [{\citenamefont {Bashinsky}\ and\ \citenamefont
  {Seljak}(2004)}]{Bashinsky:2003tk}%
  \BibitemOpen
  \bibfield  {author} {\bibinfo {author} {\bibfnamefont {S.}~\bibnamefont
  {Bashinsky}}\ and\ \bibinfo {author} {\bibfnamefont {U.}~\bibnamefont
  {Seljak}},\ }\href {\doibase 10.1103/PhysRevD.69.083002} {\bibfield
  {journal} {\bibinfo  {journal} {Phys.Rev.}\ }\textbf {\bibinfo {volume}
  {D69}},\ \bibinfo {pages} {083002} (\bibinfo {year} {2004})}\BibitemShut
  {NoStop}%
\bibitem [{\citenamefont {{Hou}}\ \emph {et~al.}(2013)\citenamefont {{Hou}},
  \citenamefont {{Keisler}}, \citenamefont {{Knox}}, \citenamefont {{Millea}},\
  and\ \citenamefont {{Reichardt}}}]{Hou:2011ec}%
  \BibitemOpen
  \bibfield  {author} {\bibinfo {author} {\bibfnamefont {Z.}~\bibnamefont
  {{Hou}}}, \bibinfo {author} {\bibfnamefont {R.}~\bibnamefont {{Keisler}}},
  \bibinfo {author} {\bibfnamefont {L.}~\bibnamefont {{Knox}}}, \bibinfo
  {author} {\bibfnamefont {M.}~\bibnamefont {{Millea}}}, \ and\ \bibinfo
  {author} {\bibfnamefont {C.}~\bibnamefont {{Reichardt}}},\ }\href {\doibase
  10.1103/PhysRevD.87.083008} {\bibfield  {journal} {\bibinfo  {journal}
  {\prd}\ }\textbf {\bibinfo {volume} {87}},\ \bibinfo {eid} {083008} (\bibinfo
  {year} {2013})},\ \Eprint {http://arxiv.org/abs/1104.2333} {arXiv:1104.2333
  [astro-ph.CO]} \BibitemShut {NoStop}%
\bibitem [{\citenamefont {Abe}\ \emph {et~al.}(2013)\citenamefont {Abe} \emph
  {et~al.}}]{Abe:2012av}%
  \BibitemOpen
  \bibfield  {author} {\bibinfo {author} {\bibfnamefont {K.}~\bibnamefont
  {Abe}} \emph {et~al.} (\bibinfo {collaboration} {T2K Collaboration}),\ }\href
  {\doibase 10.1103/PhysRevD.87.012001, 10.1103/PhysRevD.87.019902} {\bibfield
  {journal} {\bibinfo  {journal} {Phys.Rev.}\ }\textbf {\bibinfo {volume}
  {D87}},\ \bibinfo {pages} {012001} (\bibinfo {year} {2013})}\BibitemShut
  {NoStop}%
\bibitem [{\citenamefont {Trotta}\ and\ \citenamefont
  {Melchiorri}(2005)}]{Trotta:2004ty}%
  \BibitemOpen
  \bibfield  {author} {\bibinfo {author} {\bibfnamefont {R.}~\bibnamefont
  {Trotta}}\ and\ \bibinfo {author} {\bibfnamefont {A.}~\bibnamefont
  {Melchiorri}},\ }\href {\doibase 10.1103/PhysRevLett.95.011305} {\bibfield
  {journal} {\bibinfo  {journal} {Phys.Rev.Lett.}\ }\textbf {\bibinfo {volume}
  {95}},\ \bibinfo {pages} {011305} (\bibinfo {year} {2005})}\BibitemShut
  {NoStop}%
\bibitem [{\citenamefont {Melchiorri}\ and\ \citenamefont
  {Serra}(2006)}]{Melchiorri:2006xs}%
  \BibitemOpen
  \bibfield  {author} {\bibinfo {author} {\bibfnamefont {A.}~\bibnamefont
  {Melchiorri}}\ and\ \bibinfo {author} {\bibfnamefont {P.}~\bibnamefont
  {Serra}},\ }\href {\doibase 10.1103/PhysRevD.74.127301} {\bibfield  {journal}
  {\bibinfo  {journal} {Phys.Rev.}\ }\textbf {\bibinfo {volume} {D74}},\
  \bibinfo {pages} {127301} (\bibinfo {year} {2006})}\BibitemShut {NoStop}%
\bibitem [{\citenamefont {De~Bernardis}\ \emph {et~al.}(2008)\citenamefont
  {De~Bernardis}, \citenamefont {Pagano}, \citenamefont {Serra}, \citenamefont
  {Melchiorri} \emph {et~al.}}]{DeBernardis:2008ys}%
  \BibitemOpen
  \bibfield  {author} {\bibinfo {author} {\bibfnamefont {F.}~\bibnamefont
  {De~Bernardis}}, \bibinfo {author} {\bibfnamefont {L.}~\bibnamefont
  {Pagano}}, \bibinfo {author} {\bibfnamefont {P.}~\bibnamefont {Serra}},
  \bibinfo {author} {\bibfnamefont {A.}~\bibnamefont {Melchiorri}},  \emph
  {et~al.},\ }\href {\doibase 10.1088/1475-7516/2008/06/013} {\bibfield
  {journal} {\bibinfo  {journal} {JCAP}\ }\textbf {\bibinfo {volume} {0806}},\
  \bibinfo {pages} {013} (\bibinfo {year} {2008})}\BibitemShut {NoStop}%
\bibitem [{\citenamefont {Smith}\ \emph {et~al.}(2012)\citenamefont {Smith},
  \citenamefont {Das},\ and\ \citenamefont {Zahn}}]{Smith:2011es}%
  \BibitemOpen
  \bibfield  {author} {\bibinfo {author} {\bibfnamefont {T.~L.}\ \bibnamefont
  {Smith}}, \bibinfo {author} {\bibfnamefont {S.}~\bibnamefont {Das}}, \ and\
  \bibinfo {author} {\bibfnamefont {O.}~\bibnamefont {Zahn}},\ }\href {\doibase
  10.1103/PhysRevD.85.023001} {\bibfield  {journal} {\bibinfo  {journal}
  {Phys.Rev.}\ }\textbf {\bibinfo {volume} {D85}},\ \bibinfo {pages} {023001}
  (\bibinfo {year} {2012})}\BibitemShut {NoStop}%
\bibitem [{\citenamefont {Archidiacono}\ \emph {et~al.}(2011)\citenamefont
  {Archidiacono}, \citenamefont {Calabrese},\ and\ \citenamefont
  {Melchiorri}}]{Archidiacono:2011gq}%
  \BibitemOpen
  \bibfield  {author} {\bibinfo {author} {\bibfnamefont {M.}~\bibnamefont
  {Archidiacono}}, \bibinfo {author} {\bibfnamefont {E.}~\bibnamefont
  {Calabrese}}, \ and\ \bibinfo {author} {\bibfnamefont {A.}~\bibnamefont
  {Melchiorri}},\ }\href {\doibase 10.1103/PhysRevD.84.123008} {\bibfield
  {journal} {\bibinfo  {journal} {Phys.Rev.}\ }\textbf {\bibinfo {volume}
  {D84}},\ \bibinfo {pages} {123008} (\bibinfo {year} {2011})}\BibitemShut
  {NoStop}%
\bibitem [{\citenamefont {Archidiacono}\ \emph {et~al.}(2012)\citenamefont
  {Archidiacono}, \citenamefont {Giusarma}, \citenamefont {Melchiorri},\ and\
  \citenamefont {Mena}}]{Archidiacono:2012gv}%
  \BibitemOpen
  \bibfield  {author} {\bibinfo {author} {\bibfnamefont {M.}~\bibnamefont
  {Archidiacono}}, \bibinfo {author} {\bibfnamefont {E.}~\bibnamefont
  {Giusarma}}, \bibinfo {author} {\bibfnamefont {A.}~\bibnamefont
  {Melchiorri}}, \ and\ \bibinfo {author} {\bibfnamefont {O.}~\bibnamefont
  {Mena}},\ }\href {\doibase 10.1103/PhysRevD.86.043509} {\bibfield  {journal}
  {\bibinfo  {journal} {Phys.Rev.}\ }\textbf {\bibinfo {volume} {D86}},\
  \bibinfo {pages} {043509} (\bibinfo {year} {2012})}\BibitemShut {NoStop}%
\bibitem [{\citenamefont {Gerbino}\ \emph {et~al.}(2013)\citenamefont
  {Gerbino}, \citenamefont {Di~Valentino},\ and\ \citenamefont
  {Said}}]{Gerbino:2013ova}%
  \BibitemOpen
  \bibfield  {author} {\bibinfo {author} {\bibfnamefont {M.}~\bibnamefont
  {Gerbino}}, \bibinfo {author} {\bibfnamefont {E.}~\bibnamefont
  {Di~Valentino}}, \ and\ \bibinfo {author} {\bibfnamefont {N.}~\bibnamefont
  {Said}},\ }\href {\doibase 10.1103/PhysRevD.88.063538} {\bibfield  {journal}
  {\bibinfo  {journal} {Phys.Rev.}\ }\textbf {\bibinfo {volume} {D88}},\
  \bibinfo {pages} {063538} (\bibinfo {year} {2013})},\ \Eprint
  {http://arxiv.org/abs/1304.7400} {arXiv:1304.7400 [astro-ph.CO]} \BibitemShut
  {NoStop}%
\bibitem [{\citenamefont {Hu}(1998)}]{Hu:1998kj}%
  \BibitemOpen
  \bibfield  {author} {\bibinfo {author} {\bibfnamefont {W.}~\bibnamefont
  {Hu}},\ }\href {\doibase 10.1086/306274} {\bibfield  {journal} {\bibinfo
  {journal} {Astrophys.J.}\ }\textbf {\bibinfo {volume} {506}},\ \bibinfo
  {pages} {485} (\bibinfo {year} {1998})}\BibitemShut {NoStop}%
\bibitem [{\citenamefont {Gelmini}\ and\ \citenamefont
  {Roncadelli}(1981)}]{Gelmini:1980re}%
  \BibitemOpen
  \bibfield  {author} {\bibinfo {author} {\bibfnamefont {G.}~\bibnamefont
  {Gelmini}}\ and\ \bibinfo {author} {\bibfnamefont {M.}~\bibnamefont
  {Roncadelli}},\ }\href {\doibase 10.1016/0370-2693(81)90559-1} {\bibfield
  {journal} {\bibinfo  {journal} {Phys.Lett.}\ }\textbf {\bibinfo {volume}
  {B99}},\ \bibinfo {pages} {411} (\bibinfo {year} {1981})}\BibitemShut
  {NoStop}%
\bibitem [{\citenamefont {Chikashige}\ \emph {et~al.}(1980)\citenamefont
  {Chikashige}, \citenamefont {Mohapatra},\ and\ \citenamefont
  {Peccei}}]{Chikashige:1980qk}%
  \BibitemOpen
  \bibfield  {author} {\bibinfo {author} {\bibfnamefont {Y.}~\bibnamefont
  {Chikashige}}, \bibinfo {author} {\bibfnamefont {R.~N.}\ \bibnamefont
  {Mohapatra}}, \ and\ \bibinfo {author} {\bibfnamefont {R.}~\bibnamefont
  {Peccei}},\ }\href {\doibase 10.1103/PhysRevLett.45.1926} {\bibfield
  {journal} {\bibinfo  {journal} {Phys.Rev.Lett.}\ }\textbf {\bibinfo {volume}
  {45}},\ \bibinfo {pages} {1926} (\bibinfo {year} {1980})}\BibitemShut
  {NoStop}%
\bibitem [{\citenamefont {Bardin}\ \emph {et~al.}(1970)\citenamefont {Bardin},
  \citenamefont {Bilenky},\ and\ \citenamefont {Pontecorvo}}]{Bardin:1970wq}%
  \BibitemOpen
  \bibfield  {author} {\bibinfo {author} {\bibfnamefont {D.~Y.}\ \bibnamefont
  {Bardin}}, \bibinfo {author} {\bibfnamefont {S.~M.}\ \bibnamefont {Bilenky}},
  \ and\ \bibinfo {author} {\bibfnamefont {B.}~\bibnamefont {Pontecorvo}},\
  }\href {\doibase 10.1016/0370-2693(70)90602-7} {\bibfield  {journal}
  {\bibinfo  {journal} {Phys.Lett.}\ }\textbf {\bibinfo {volume} {B32}},\
  \bibinfo {pages} {121} (\bibinfo {year} {1970})}\BibitemShut {NoStop}%
\bibitem [{\citenamefont {Barger}\ \emph {et~al.}(1982)\citenamefont {Barger},
  \citenamefont {Keung},\ and\ \citenamefont {Pakvasa}}]{Barger:1981vd}%
  \BibitemOpen
  \bibfield  {author} {\bibinfo {author} {\bibfnamefont {V.~D.}\ \bibnamefont
  {Barger}}, \bibinfo {author} {\bibfnamefont {W.-Y.}\ \bibnamefont {Keung}}, \
  and\ \bibinfo {author} {\bibfnamefont {S.}~\bibnamefont {Pakvasa}},\ }\href
  {\doibase 10.1103/PhysRevD.25.907} {\bibfield  {journal} {\bibinfo  {journal}
  {Phys.Rev.}\ }\textbf {\bibinfo {volume} {D25}},\ \bibinfo {pages} {907}
  (\bibinfo {year} {1982})}\BibitemShut {NoStop}%
\bibitem [{\citenamefont {Lessa}\ and\ \citenamefont
  {Peres}(2007)}]{Lessa:2007up}%
  \BibitemOpen
  \bibfield  {author} {\bibinfo {author} {\bibfnamefont {A.~P.}\ \bibnamefont
  {Lessa}}\ and\ \bibinfo {author} {\bibfnamefont {O.~L.~G.}\ \bibnamefont
  {Peres}},\ }\href {\doibase 10.1103/PhysRevD.75.094001} {\bibfield  {journal}
  {\bibinfo  {journal} {Phys.Rev.}\ }\textbf {\bibinfo {volume} {D75}},\
  \bibinfo {pages} {094001} (\bibinfo {year} {2007})}\BibitemShut {NoStop}%
\bibitem [{\citenamefont {Bilenky}\ \emph {et~al.}(1993)\citenamefont
  {Bilenky}, \citenamefont {Bilenky},\ and\ \citenamefont
  {Santamaria}}]{Bilenky:1992xn}%
  \BibitemOpen
  \bibfield  {author} {\bibinfo {author} {\bibfnamefont {M.~S.}\ \bibnamefont
  {Bilenky}}, \bibinfo {author} {\bibfnamefont {S.~M.}\ \bibnamefont
  {Bilenky}}, \ and\ \bibinfo {author} {\bibfnamefont {A.}~\bibnamefont
  {Santamaria}},\ }\href {\doibase 10.1016/0370-2693(93)90703-K} {\bibfield
  {journal} {\bibinfo  {journal} {Phys.Lett.}\ }\textbf {\bibinfo {volume}
  {B301}},\ \bibinfo {pages} {287} (\bibinfo {year} {1993})}\BibitemShut
  {NoStop}%
\bibitem [{\citenamefont {Bilenky}\ and\ \citenamefont
  {Santamaria}(1999)}]{Bilenky:1999dn}%
  \BibitemOpen
  \bibfield  {author} {\bibinfo {author} {\bibfnamefont {M.~S.}\ \bibnamefont
  {Bilenky}}\ and\ \bibinfo {author} {\bibfnamefont {A.}~\bibnamefont
  {Santamaria}},\ }\href@noop {} {\  (\bibinfo {year} {1999})},\ \Eprint
  {http://arxiv.org/abs/hep-ph/9908272} {arXiv:hep-ph/9908272 [hep-ph]}
  \BibitemShut {NoStop}%
\bibitem [{\citenamefont {Laha}\ \emph {et~al.}(2014)\citenamefont {Laha},
  \citenamefont {Dasgupta},\ and\ \citenamefont {Beacom}}]{Laha:2013xua}%
  \BibitemOpen
  \bibfield  {author} {\bibinfo {author} {\bibfnamefont {R.}~\bibnamefont
  {Laha}}, \bibinfo {author} {\bibfnamefont {B.}~\bibnamefont {Dasgupta}}, \
  and\ \bibinfo {author} {\bibfnamefont {J.~F.}\ \bibnamefont {Beacom}},\
  }\href {\doibase 10.1103/PhysRevD.89.093025} {\bibfield  {journal} {\bibinfo
  {journal} {Phys.Rev.}\ }\textbf {\bibinfo {volume} {D89}},\ \bibinfo {pages}
  {093025} (\bibinfo {year} {2014})},\ \Eprint {http://arxiv.org/abs/1304.3460}
  {arXiv:1304.3460 [hep-ph]} \BibitemShut {NoStop}%
\bibitem [{\citenamefont {Ahlgren}\ \emph {et~al.}(2013)\citenamefont
  {Ahlgren}, \citenamefont {Ohlsson},\ and\ \citenamefont
  {Zhou}}]{Ahlgren:2013wba}%
  \BibitemOpen
  \bibfield  {author} {\bibinfo {author} {\bibfnamefont {B.}~\bibnamefont
  {Ahlgren}}, \bibinfo {author} {\bibfnamefont {T.}~\bibnamefont {Ohlsson}}, \
  and\ \bibinfo {author} {\bibfnamefont {S.}~\bibnamefont {Zhou}},\ }\href
  {\doibase 10.1103/PhysRevLett.111.199001} {\bibfield  {journal} {\bibinfo
  {journal} {Phys.Rev.Lett.}\ }\textbf {\bibinfo {volume} {111}},\ \bibinfo
  {pages} {199001} (\bibinfo {year} {2013})},\ \Eprint
  {http://arxiv.org/abs/1309.0991} {arXiv:1309.0991 [hep-ph]} \BibitemShut
  {NoStop}%
\bibitem [{\citenamefont {Choi}\ and\ \citenamefont
  {Santamaria}(1990)}]{Choi:1989hi}%
  \BibitemOpen
  \bibfield  {author} {\bibinfo {author} {\bibfnamefont {K.}~\bibnamefont
  {Choi}}\ and\ \bibinfo {author} {\bibfnamefont {A.}~\bibnamefont
  {Santamaria}},\ }\href {\doibase 10.1103/PhysRevD.42.293} {\bibfield
  {journal} {\bibinfo  {journal} {Phys.Rev.}\ }\textbf {\bibinfo {volume}
  {D42}},\ \bibinfo {pages} {293} (\bibinfo {year} {1990})}\BibitemShut
  {NoStop}%
\bibitem [{\citenamefont {Kachelriess}\ \emph {et~al.}(2000)\citenamefont
  {Kachelriess}, \citenamefont {Tomas},\ and\ \citenamefont
  {Valle}}]{Kachelriess:2000qc}%
  \BibitemOpen
  \bibfield  {author} {\bibinfo {author} {\bibfnamefont {M.}~\bibnamefont
  {Kachelriess}}, \bibinfo {author} {\bibfnamefont {R.}~\bibnamefont {Tomas}},
  \ and\ \bibinfo {author} {\bibfnamefont {J.}~\bibnamefont {Valle}},\ }\href
  {\doibase 10.1103/PhysRevD.62.023004} {\bibfield  {journal} {\bibinfo
  {journal} {Phys.Rev.}\ }\textbf {\bibinfo {volume} {D62}},\ \bibinfo {pages}
  {023004} (\bibinfo {year} {2000})},\ \Eprint
  {http://arxiv.org/abs/hep-ph/0001039} {arXiv:hep-ph/0001039 [hep-ph]}
  \BibitemShut {NoStop}%
\bibitem [{\citenamefont {Farzan}(2003)}]{Farzan:2002wx}%
  \BibitemOpen
  \bibfield  {author} {\bibinfo {author} {\bibfnamefont {Y.}~\bibnamefont
  {Farzan}},\ }\href {\doibase 10.1103/PhysRevD.67.073015} {\bibfield
  {journal} {\bibinfo  {journal} {Phys.Rev.}\ }\textbf {\bibinfo {volume}
  {D67}},\ \bibinfo {pages} {073015} (\bibinfo {year} {2003})},\ \Eprint
  {http://arxiv.org/abs/hep-ph/0211375} {arXiv:hep-ph/0211375 [hep-ph]}
  \BibitemShut {NoStop}%
\bibitem [{\citenamefont {Davoudiasl}\ and\ \citenamefont
  {Huber}(2005)}]{Davoudiasl:2005fd}%
  \BibitemOpen
  \bibfield  {author} {\bibinfo {author} {\bibfnamefont {H.}~\bibnamefont
  {Davoudiasl}}\ and\ \bibinfo {author} {\bibfnamefont {P.}~\bibnamefont
  {Huber}},\ }\href {\doibase 10.1103/PhysRevLett.95.191302} {\bibfield
  {journal} {\bibinfo  {journal} {Phys.Rev.Lett.}\ }\textbf {\bibinfo {volume}
  {95}},\ \bibinfo {pages} {191302} (\bibinfo {year} {2005})},\ \Eprint
  {http://arxiv.org/abs/hep-ph/0504265} {arXiv:hep-ph/0504265 [hep-ph]}
  \BibitemShut {NoStop}%
\bibitem [{\citenamefont {Fayet}\ \emph {et~al.}(2006)\citenamefont {Fayet},
  \citenamefont {Hooper},\ and\ \citenamefont {Sigl}}]{Fayet:2006sa}%
  \BibitemOpen
  \bibfield  {author} {\bibinfo {author} {\bibfnamefont {P.}~\bibnamefont
  {Fayet}}, \bibinfo {author} {\bibfnamefont {D.}~\bibnamefont {Hooper}}, \
  and\ \bibinfo {author} {\bibfnamefont {G.}~\bibnamefont {Sigl}},\ }\href
  {\doibase 10.1103/PhysRevLett.96.211302} {\bibfield  {journal} {\bibinfo
  {journal} {Phys.Rev.Lett.}\ }\textbf {\bibinfo {volume} {96}},\ \bibinfo
  {pages} {211302} (\bibinfo {year} {2006})},\ \Eprint
  {http://arxiv.org/abs/hep-ph/0602169} {arXiv:hep-ph/0602169 [hep-ph]}
  \BibitemShut {NoStop}%
\bibitem [{\citenamefont {Blennow}\ \emph {et~al.}(2008)\citenamefont
  {Blennow}, \citenamefont {Mirizzi},\ and\ \citenamefont
  {Serpico}}]{Blennow:2008er}%
  \BibitemOpen
  \bibfield  {author} {\bibinfo {author} {\bibfnamefont {M.}~\bibnamefont
  {Blennow}}, \bibinfo {author} {\bibfnamefont {A.}~\bibnamefont {Mirizzi}}, \
  and\ \bibinfo {author} {\bibfnamefont {P.~D.}\ \bibnamefont {Serpico}},\
  }\href {\doibase 10.1103/PhysRevD.78.113004} {\bibfield  {journal} {\bibinfo
  {journal} {Phys.Rev.}\ }\textbf {\bibinfo {volume} {D78}},\ \bibinfo {pages}
  {113004} (\bibinfo {year} {2008})},\ \Eprint {http://arxiv.org/abs/0810.2297}
  {arXiv:0810.2297 [hep-ph]} \BibitemShut {NoStop}%
\bibitem [{\citenamefont {Sher}\ and\ \citenamefont
  {Triola}(2011)}]{Sher:2011mx}%
  \BibitemOpen
  \bibfield  {author} {\bibinfo {author} {\bibfnamefont {M.}~\bibnamefont
  {Sher}}\ and\ \bibinfo {author} {\bibfnamefont {C.}~\bibnamefont {Triola}},\
  }\href {\doibase 10.1103/PhysRevD.83.117702} {\bibfield  {journal} {\bibinfo
  {journal} {Phys.Rev.}\ }\textbf {\bibinfo {volume} {D83}},\ \bibinfo {pages}
  {117702} (\bibinfo {year} {2011})},\ \Eprint {http://arxiv.org/abs/1105.4844}
  {arXiv:1105.4844 [hep-ph]} \BibitemShut {NoStop}%
\bibitem [{\citenamefont {Zhou}(2011)}]{Zhou:2011rc}%
  \BibitemOpen
  \bibfield  {author} {\bibinfo {author} {\bibfnamefont {S.}~\bibnamefont
  {Zhou}},\ }\href {\doibase 10.1103/PhysRevD.84.038701} {\bibfield  {journal}
  {\bibinfo  {journal} {Phys.Rev.}\ }\textbf {\bibinfo {volume} {D84}},\
  \bibinfo {pages} {038701} (\bibinfo {year} {2011})},\ \Eprint
  {http://arxiv.org/abs/1106.3880} {arXiv:1106.3880 [hep-ph]} \BibitemShut
  {NoStop}%
\bibitem [{\citenamefont {Manohar}(1987)}]{Manohar:1987ec}%
  \BibitemOpen
  \bibfield  {author} {\bibinfo {author} {\bibfnamefont {A.}~\bibnamefont
  {Manohar}},\ }\href {\doibase 10.1016/0370-2693(87)91171-3} {\bibfield
  {journal} {\bibinfo  {journal} {Phys.Lett.}\ }\textbf {\bibinfo {volume}
  {B192}},\ \bibinfo {pages} {217} (\bibinfo {year} {1987})}\BibitemShut
  {NoStop}%
\bibitem [{\citenamefont {Dicus}\ \emph {et~al.}(1989)\citenamefont {Dicus},
  \citenamefont {Nussinov}, \citenamefont {Pal},\ and\ \citenamefont
  {Teplitz}}]{Dicus:1988jh}%
  \BibitemOpen
  \bibfield  {author} {\bibinfo {author} {\bibfnamefont {D.~A.}\ \bibnamefont
  {Dicus}}, \bibinfo {author} {\bibfnamefont {S.}~\bibnamefont {Nussinov}},
  \bibinfo {author} {\bibfnamefont {P.~B.}\ \bibnamefont {Pal}}, \ and\
  \bibinfo {author} {\bibfnamefont {V.~L.}\ \bibnamefont {Teplitz}},\ }\href
  {\doibase 10.1016/0370-2693(89)90480-2} {\bibfield  {journal} {\bibinfo
  {journal} {Phys.Lett.}\ }\textbf {\bibinfo {volume} {B218}},\ \bibinfo
  {pages} {84} (\bibinfo {year} {1989})}\BibitemShut {NoStop}%
\bibitem [{\citenamefont {Kolb}\ and\ \citenamefont
  {Turner}(1987)}]{Kolb:1987qy}%
  \BibitemOpen
  \bibfield  {author} {\bibinfo {author} {\bibfnamefont {E.~W.}\ \bibnamefont
  {Kolb}}\ and\ \bibinfo {author} {\bibfnamefont {M.~S.}\ \bibnamefont
  {Turner}},\ }\href {\doibase 10.1103/PhysRevD.36.2895} {\bibfield  {journal}
  {\bibinfo  {journal} {Phys.Rev.}\ }\textbf {\bibinfo {volume} {D36}},\
  \bibinfo {pages} {2895} (\bibinfo {year} {1987})}\BibitemShut {NoStop}%
\bibitem [{\citenamefont {Keranen}(1998)}]{Keranen:1997gz}%
  \BibitemOpen
  \bibfield  {author} {\bibinfo {author} {\bibfnamefont {P.}~\bibnamefont
  {Keranen}},\ }\href {\doibase 10.1016/S0370-2693(97)01405-6} {\bibfield
  {journal} {\bibinfo  {journal} {Phys.Lett.}\ }\textbf {\bibinfo {volume}
  {B417}},\ \bibinfo {pages} {320} (\bibinfo {year} {1998})}\BibitemShut
  {NoStop}%
\bibitem [{\citenamefont {Hooper}(2007)}]{Hooper:2007jr}%
  \BibitemOpen
  \bibfield  {author} {\bibinfo {author} {\bibfnamefont {D.}~\bibnamefont
  {Hooper}},\ }\href {\doibase 10.1103/PhysRevD.75.123001} {\bibfield
  {journal} {\bibinfo  {journal} {Phys.Rev.}\ }\textbf {\bibinfo {volume}
  {D75}},\ \bibinfo {pages} {123001} (\bibinfo {year} {2007})},\ \Eprint
  {http://arxiv.org/abs/hep-ph/0701194} {arXiv:hep-ph/0701194 [hep-ph]}
  \BibitemShut {NoStop}%
\bibitem [{\citenamefont {Ng}\ and\ \citenamefont {Beacom}(2014)}]{Ng:2014pca}%
  \BibitemOpen
  \bibfield  {author} {\bibinfo {author} {\bibfnamefont {K.~C.~Y.}\
  \bibnamefont {Ng}}\ and\ \bibinfo {author} {\bibfnamefont {J.~F.}\
  \bibnamefont {Beacom}},\ }\href {\doibase 10.1103/PhysRevD.90.065035}
  {\bibfield  {journal} {\bibinfo  {journal} {Phys.Rev.}\ }\textbf {\bibinfo
  {volume} {D90}},\ \bibinfo {pages} {065035} (\bibinfo {year} {2014})},\
  \Eprint {http://arxiv.org/abs/1404.2288} {arXiv:1404.2288 [astro-ph.HE]}
  \BibitemShut {NoStop}%
\bibitem [{\citenamefont {Ioka}\ and\ \citenamefont
  {Murase}(2014)}]{Ioka:2014kca}%
  \BibitemOpen
  \bibfield  {author} {\bibinfo {author} {\bibfnamefont {K.}~\bibnamefont
  {Ioka}}\ and\ \bibinfo {author} {\bibfnamefont {K.}~\bibnamefont {Murase}},\
  }\href {\doibase 10.1093/ptep/ptu090} {\bibfield  {journal} {\bibinfo
  {journal} {PTEP}\ }\textbf {\bibinfo {volume} {2014}},\ \bibinfo {pages}
  {061E01} (\bibinfo {year} {2014})},\ \Eprint {http://arxiv.org/abs/1404.2279}
  {arXiv:1404.2279 [astro-ph.HE]} \BibitemShut {NoStop}%
\bibitem [{\citenamefont {Cherry}\ \emph {et~al.}(2014)\citenamefont {Cherry},
  \citenamefont {Friedland},\ and\ \citenamefont {Shoemaker}}]{Cherry:2014xra}%
  \BibitemOpen
  \bibfield  {author} {\bibinfo {author} {\bibfnamefont {J.~F.}\ \bibnamefont
  {Cherry}}, \bibinfo {author} {\bibfnamefont {A.}~\bibnamefont {Friedland}}, \
  and\ \bibinfo {author} {\bibfnamefont {I.~M.}\ \bibnamefont {Shoemaker}},\
  }\href@noop {} {\  (\bibinfo {year} {2014})},\ \Eprint
  {http://arxiv.org/abs/1411.1071} {arXiv:1411.1071 [hep-ph]} \BibitemShut
  {NoStop}%
\bibitem [{\citenamefont {Boehm}\ \emph {et~al.}(2012)\citenamefont {Boehm},
  \citenamefont {Dolan},\ and\ \citenamefont {McCabe}}]{Boehm:2012gr}%
  \BibitemOpen
  \bibfield  {author} {\bibinfo {author} {\bibfnamefont {C.}~\bibnamefont
  {Boehm}}, \bibinfo {author} {\bibfnamefont {M.~J.}\ \bibnamefont {Dolan}}, \
  and\ \bibinfo {author} {\bibfnamefont {C.}~\bibnamefont {McCabe}},\ }\href
  {\doibase 10.1088/1475-7516/2012/12/027} {\bibfield  {journal} {\bibinfo
  {journal} {JCAP}\ }\textbf {\bibinfo {volume} {1212}},\ \bibinfo {pages}
  {027} (\bibinfo {year} {2012})}\BibitemShut {NoStop}%
\bibitem [{\citenamefont {Ma}\ and\ \citenamefont
  {Bertschinger}(1995)}]{Ma:1995ey}%
  \BibitemOpen
  \bibfield  {author} {\bibinfo {author} {\bibfnamefont {C.-P.}\ \bibnamefont
  {Ma}}\ and\ \bibinfo {author} {\bibfnamefont {E.}~\bibnamefont
  {Bertschinger}},\ }\href {\doibase 10.1086/176550} {\bibfield  {journal}
  {\bibinfo  {journal} {Astrophys.J.}\ }\textbf {\bibinfo {volume} {455}},\
  \bibinfo {pages} {7} (\bibinfo {year} {1995})}\BibitemShut {NoStop}%
\bibitem [{\citenamefont {Oldengott}\ \emph {et~al.}(2014)\citenamefont
  {Oldengott}, \citenamefont {Rampf},\ and\ \citenamefont
  {Wong}}]{Oldengott:2014qra}%
  \BibitemOpen
  \bibfield  {author} {\bibinfo {author} {\bibfnamefont {I.~M.}\ \bibnamefont
  {Oldengott}}, \bibinfo {author} {\bibfnamefont {C.}~\bibnamefont {Rampf}}, \
  and\ \bibinfo {author} {\bibfnamefont {Y.~Y.~Y.}\ \bibnamefont {Wong}},\
  }\href@noop {} {\  (\bibinfo {year} {2014})},\ \Eprint
  {http://arxiv.org/abs/1409.1577} {arXiv:1409.1577 [astro-ph.CO]} \BibitemShut
  {NoStop}%
\bibitem [{\citenamefont {Lewis}\ \emph {et~al.}(2000)\citenamefont {Lewis},
  \citenamefont {Challinor},\ and\ \citenamefont {Lasenby}}]{Lewis:1999bs}%
  \BibitemOpen
  \bibfield  {author} {\bibinfo {author} {\bibfnamefont {A.}~\bibnamefont
  {Lewis}}, \bibinfo {author} {\bibfnamefont {A.}~\bibnamefont {Challinor}}, \
  and\ \bibinfo {author} {\bibfnamefont {A.}~\bibnamefont {Lasenby}},\ }\href
  {\doibase 10.1086/309179} {\bibfield  {journal} {\bibinfo  {journal}
  {Astrophys.J.}\ }\textbf {\bibinfo {volume} {538}},\ \bibinfo {pages} {473}
  (\bibinfo {year} {2000})}\BibitemShut {NoStop}%
\bibitem [{\citenamefont {Cyr-Racine}\ and\ \citenamefont
  {Sigurdson}(2011)}]{CyrRacine:2010bk}%
  \BibitemOpen
  \bibfield  {author} {\bibinfo {author} {\bibfnamefont {F.-Y.}\ \bibnamefont
  {Cyr-Racine}}\ and\ \bibinfo {author} {\bibfnamefont {K.}~\bibnamefont
  {Sigurdson}},\ }\href {\doibase 10.1103/PhysRevD.83.103521} {\bibfield
  {journal} {\bibinfo  {journal} {Phys.Rev.}\ }\textbf {\bibinfo {volume}
  {D83}},\ \bibinfo {pages} {103521} (\bibinfo {year} {2011})}\BibitemShut
  {NoStop}%
\bibitem [{\citenamefont {Dunkley}\ \emph {et~al.}(2013)\citenamefont
  {Dunkley}, \citenamefont {Calabrese}, \citenamefont {Sievers}, \citenamefont
  {Addison}, \citenamefont {Battaglia} \emph {et~al.}}]{Dunkley:2013vu}%
  \BibitemOpen
  \bibfield  {author} {\bibinfo {author} {\bibfnamefont {J.}~\bibnamefont
  {Dunkley}}, \bibinfo {author} {\bibfnamefont {E.}~\bibnamefont {Calabrese}},
  \bibinfo {author} {\bibfnamefont {J.}~\bibnamefont {Sievers}}, \bibinfo
  {author} {\bibfnamefont {G.}~\bibnamefont {Addison}}, \bibinfo {author}
  {\bibfnamefont {N.}~\bibnamefont {Battaglia}},  \emph {et~al.},\ }\href
  {\doibase 10.1088/1475-7516/2013/07/025} {\bibfield  {journal} {\bibinfo
  {journal} {JCAP}\ }\textbf {\bibinfo {volume} {1307}},\ \bibinfo {pages}
  {025} (\bibinfo {year} {2013})},\ \Eprint {http://arxiv.org/abs/1301.0776}
  {arXiv:1301.0776 [astro-ph.CO]} \BibitemShut {NoStop}%
\bibitem [{\citenamefont {Das}\ \emph {et~al.}(2014)\citenamefont {Das},
  \citenamefont {Louis}, \citenamefont {Nolta}, \citenamefont {Addison},
  \citenamefont {Battistelli} \emph {et~al.}}]{Das:2013zf}%
  \BibitemOpen
  \bibfield  {author} {\bibinfo {author} {\bibfnamefont {S.}~\bibnamefont
  {Das}}, \bibinfo {author} {\bibfnamefont {T.}~\bibnamefont {Louis}}, \bibinfo
  {author} {\bibfnamefont {M.~R.}\ \bibnamefont {Nolta}}, \bibinfo {author}
  {\bibfnamefont {G.~E.}\ \bibnamefont {Addison}}, \bibinfo {author}
  {\bibfnamefont {E.~S.}\ \bibnamefont {Battistelli}},  \emph {et~al.},\ }\href
  {\doibase 10.1088/1475-7516/2014/04/014} {\bibfield  {journal} {\bibinfo
  {journal} {JCAP}\ }\textbf {\bibinfo {volume} {1404}},\ \bibinfo {pages}
  {014} (\bibinfo {year} {2014})},\ \Eprint {http://arxiv.org/abs/1301.1037}
  {arXiv:1301.1037 [astro-ph.CO]} \BibitemShut {NoStop}%
\bibitem [{\citenamefont {{Story}}\ \emph {et~al.}(2013)\citenamefont
  {{Story}}, \citenamefont {{Reichardt}}, \citenamefont {{Hou}}, \citenamefont
  {{Keisler}}, \citenamefont {{Aird}}, \citenamefont {{Benson}}, \citenamefont
  {{Bleem}}, \citenamefont {{Carlstrom}}, \citenamefont {{Chang}},
  \citenamefont {{Cho}}, \citenamefont {{Crawford}}, \citenamefont {{Crites}},
  \citenamefont {{de Haan}}, \citenamefont {{Dobbs}}, \citenamefont {{Dudley}},
  \citenamefont {{Follin}}, \citenamefont {{George}}, \citenamefont
  {{Halverson}}, \citenamefont {{Holder}}, \citenamefont {{Holzapfel}},
  \citenamefont {{Hoover}}, \citenamefont {{Hrubes}}, \citenamefont {{Joy}},
  \citenamefont {{Knox}}, \citenamefont {{Lee}}, \citenamefont {{Leitch}},
  \citenamefont {{Lueker}}, \citenamefont {{Luong-Van}}, \citenamefont
  {{McMahon}}, \citenamefont {{Mehl}}, \citenamefont {{Meyer}}, \citenamefont
  {{Millea}}, \citenamefont {{Mohr}}, \citenamefont {{Montroy}}, \citenamefont
  {{Padin}}, \citenamefont {{Plagge}}, \citenamefont {{Pryke}}, \citenamefont
  {{Ruhl}}, \citenamefont {{Sayre}}, \citenamefont {{Schaffer}}, \citenamefont
  {{Shaw}}, \citenamefont {{Shirokoff}}, \citenamefont {{Spieler}},
  \citenamefont {{Staniszewski}}, \citenamefont {{Stark}}, \citenamefont {{van
  Engelen}}, \citenamefont {{Vanderlinde}}, \citenamefont {{Vieira}},
  \citenamefont {{Williamson}},\ and\ \citenamefont {{Zahn}}}]{Story:2012wx}%
  \BibitemOpen
  \bibfield  {author} {\bibinfo {author} {\bibfnamefont {K.~T.}\ \bibnamefont
  {{Story}}}, \bibinfo {author} {\bibfnamefont {C.~L.}\ \bibnamefont
  {{Reichardt}}}, \bibinfo {author} {\bibfnamefont {Z.}~\bibnamefont {{Hou}}},
  \bibinfo {author} {\bibfnamefont {R.}~\bibnamefont {{Keisler}}}, \bibinfo
  {author} {\bibfnamefont {K.~A.}\ \bibnamefont {{Aird}}}, \bibinfo {author}
  {\bibfnamefont {B.~A.}\ \bibnamefont {{Benson}}}, \bibinfo {author}
  {\bibfnamefont {L.~E.}\ \bibnamefont {{Bleem}}}, \bibinfo {author}
  {\bibfnamefont {J.~E.}\ \bibnamefont {{Carlstrom}}}, \bibinfo {author}
  {\bibfnamefont {C.~L.}\ \bibnamefont {{Chang}}}, \bibinfo {author}
  {\bibfnamefont {H.-M.}\ \bibnamefont {{Cho}}}, \bibinfo {author}
  {\bibfnamefont {T.~M.}\ \bibnamefont {{Crawford}}}, \bibinfo {author}
  {\bibfnamefont {A.~T.}\ \bibnamefont {{Crites}}}, \bibinfo {author}
  {\bibfnamefont {T.}~\bibnamefont {{de Haan}}}, \bibinfo {author}
  {\bibfnamefont {M.~A.}\ \bibnamefont {{Dobbs}}}, \bibinfo {author}
  {\bibfnamefont {J.}~\bibnamefont {{Dudley}}}, \bibinfo {author}
  {\bibfnamefont {B.}~\bibnamefont {{Follin}}}, \bibinfo {author}
  {\bibfnamefont {E.~M.}\ \bibnamefont {{George}}}, \bibinfo {author}
  {\bibfnamefont {N.~W.}\ \bibnamefont {{Halverson}}}, \bibinfo {author}
  {\bibfnamefont {G.~P.}\ \bibnamefont {{Holder}}}, \bibinfo {author}
  {\bibfnamefont {W.~L.}\ \bibnamefont {{Holzapfel}}}, \bibinfo {author}
  {\bibfnamefont {S.}~\bibnamefont {{Hoover}}}, \bibinfo {author}
  {\bibfnamefont {J.~D.}\ \bibnamefont {{Hrubes}}}, \bibinfo {author}
  {\bibfnamefont {M.}~\bibnamefont {{Joy}}}, \bibinfo {author} {\bibfnamefont
  {L.}~\bibnamefont {{Knox}}}, \bibinfo {author} {\bibfnamefont {A.~T.}\
  \bibnamefont {{Lee}}}, \bibinfo {author} {\bibfnamefont {E.~M.}\ \bibnamefont
  {{Leitch}}}, \bibinfo {author} {\bibfnamefont {M.}~\bibnamefont {{Lueker}}},
  \bibinfo {author} {\bibfnamefont {D.}~\bibnamefont {{Luong-Van}}}, \bibinfo
  {author} {\bibfnamefont {J.~J.}\ \bibnamefont {{McMahon}}}, \bibinfo {author}
  {\bibfnamefont {J.}~\bibnamefont {{Mehl}}}, \bibinfo {author} {\bibfnamefont
  {S.~S.}\ \bibnamefont {{Meyer}}}, \bibinfo {author} {\bibfnamefont
  {M.}~\bibnamefont {{Millea}}}, \bibinfo {author} {\bibfnamefont {J.~J.}\
  \bibnamefont {{Mohr}}}, \bibinfo {author} {\bibfnamefont {T.~E.}\
  \bibnamefont {{Montroy}}}, \bibinfo {author} {\bibfnamefont {S.}~\bibnamefont
  {{Padin}}}, \bibinfo {author} {\bibfnamefont {T.}~\bibnamefont {{Plagge}}},
  \bibinfo {author} {\bibfnamefont {C.}~\bibnamefont {{Pryke}}}, \bibinfo
  {author} {\bibfnamefont {J.~E.}\ \bibnamefont {{Ruhl}}}, \bibinfo {author}
  {\bibfnamefont {J.~T.}\ \bibnamefont {{Sayre}}}, \bibinfo {author}
  {\bibfnamefont {K.~K.}\ \bibnamefont {{Schaffer}}}, \bibinfo {author}
  {\bibfnamefont {L.}~\bibnamefont {{Shaw}}}, \bibinfo {author} {\bibfnamefont
  {E.}~\bibnamefont {{Shirokoff}}}, \bibinfo {author} {\bibfnamefont {H.~G.}\
  \bibnamefont {{Spieler}}}, \bibinfo {author} {\bibfnamefont {Z.}~\bibnamefont
  {{Staniszewski}}}, \bibinfo {author} {\bibfnamefont {A.~A.}\ \bibnamefont
  {{Stark}}}, \bibinfo {author} {\bibfnamefont {A.}~\bibnamefont {{van
  Engelen}}}, \bibinfo {author} {\bibfnamefont {K.}~\bibnamefont
  {{Vanderlinde}}}, \bibinfo {author} {\bibfnamefont {J.~D.}\ \bibnamefont
  {{Vieira}}}, \bibinfo {author} {\bibfnamefont {R.}~\bibnamefont
  {{Williamson}}}, \ and\ \bibinfo {author} {\bibfnamefont {O.}~\bibnamefont
  {{Zahn}}},\ }\href {\doibase 10.1088/0004-637X/779/1/86} {\bibfield
  {journal} {\bibinfo  {journal} {\apj}\ }\textbf {\bibinfo {volume} {779}},\
  \bibinfo {eid} {86} (\bibinfo {year} {2013})},\ \Eprint
  {http://arxiv.org/abs/1210.7231} {arXiv:1210.7231 [astro-ph.CO]} \BibitemShut
  {NoStop}%
\bibitem [{\citenamefont {{Padmanabhan}}\ \emph {et~al.}(2012)\citenamefont
  {{Padmanabhan}}, \citenamefont {{Xu}}, \citenamefont {{Eisenstein}},
  \citenamefont {{Scalzo}}, \citenamefont {{Cuesta}}, \citenamefont {{Mehta}},\
  and\ \citenamefont {{Kazin}}}]{Padmanabhan:2012hf}%
  \BibitemOpen
  \bibfield  {author} {\bibinfo {author} {\bibfnamefont {N.}~\bibnamefont
  {{Padmanabhan}}}, \bibinfo {author} {\bibfnamefont {X.}~\bibnamefont {{Xu}}},
  \bibinfo {author} {\bibfnamefont {D.~J.}\ \bibnamefont {{Eisenstein}}},
  \bibinfo {author} {\bibfnamefont {R.}~\bibnamefont {{Scalzo}}}, \bibinfo
  {author} {\bibfnamefont {A.~J.}\ \bibnamefont {{Cuesta}}}, \bibinfo {author}
  {\bibfnamefont {K.~T.}\ \bibnamefont {{Mehta}}}, \ and\ \bibinfo {author}
  {\bibfnamefont {E.}~\bibnamefont {{Kazin}}},\ }\href {\doibase
  10.1111/j.1365-2966.2012.21888.x} {\bibfield  {journal} {\bibinfo  {journal}
  {\mnras}\ }\textbf {\bibinfo {volume} {427}},\ \bibinfo {pages} {2132}
  (\bibinfo {year} {2012})},\ \Eprint {http://arxiv.org/abs/1202.0090}
  {arXiv:1202.0090 [astro-ph.CO]} \BibitemShut {NoStop}%
\bibitem [{\citenamefont {Beutler}\ \emph {et~al.}(2011)\citenamefont
  {Beutler}, \citenamefont {Blake}, \citenamefont {Colless}, \citenamefont
  {Jones} \emph {et~al.}}]{Beutler:2011hx}%
  \BibitemOpen
  \bibfield  {author} {\bibinfo {author} {\bibfnamefont {F.}~\bibnamefont
  {Beutler}}, \bibinfo {author} {\bibfnamefont {C.}~\bibnamefont {Blake}},
  \bibinfo {author} {\bibfnamefont {M.}~\bibnamefont {Colless}}, \bibinfo
  {author} {\bibfnamefont {D.~H.}\ \bibnamefont {Jones}},  \emph {et~al.},\
  }\href {\doibase 10.1111/j.1365-2966.2011.19250.x} {\bibfield  {journal}
  {\bibinfo  {journal} {Mon.Not.Roy.Astron.Soc.}\ }\textbf {\bibinfo {volume}
  {416}},\ \bibinfo {pages} {3017} (\bibinfo {year} {2011})}\BibitemShut
  {NoStop}%
\bibitem [{\citenamefont {Anderson}\ \emph {et~al.}(2013)\citenamefont
  {Anderson}, \citenamefont {Aubourg}, \citenamefont {Bailey}, \citenamefont
  {Bizyaev} \emph {et~al.}}]{Anderson:2012sa}%
  \BibitemOpen
  \bibfield  {author} {\bibinfo {author} {\bibfnamefont {L.}~\bibnamefont
  {Anderson}}, \bibinfo {author} {\bibfnamefont {E.}~\bibnamefont {Aubourg}},
  \bibinfo {author} {\bibfnamefont {S.}~\bibnamefont {Bailey}}, \bibinfo
  {author} {\bibfnamefont {D.}~\bibnamefont {Bizyaev}},  \emph {et~al.},\
  }\href {\doibase 10.1111/j.1365-2966.2012.22066.x} {\bibfield  {journal}
  {\bibinfo  {journal} {Mon.Not.Roy.Astron.Soc.}\ }\textbf {\bibinfo {volume}
  {427}},\ \bibinfo {pages} {3435} (\bibinfo {year} {2013})}\BibitemShut
  {NoStop}%
\bibitem [{\citenamefont {Lewis}\ and\ \citenamefont
  {Bridle}(2002)}]{Lewis:2002ah}%
  \BibitemOpen
  \bibfield  {author} {\bibinfo {author} {\bibfnamefont {A.}~\bibnamefont
  {Lewis}}\ and\ \bibinfo {author} {\bibfnamefont {S.}~\bibnamefont {Bridle}},\
  }\href {\doibase 10.1103/PhysRevD.66.103511} {\bibfield  {journal} {\bibinfo
  {journal} {Phys.Rev.}\ }\textbf {\bibinfo {volume} {D66}},\ \bibinfo {pages}
  {103511} (\bibinfo {year} {2002})}\BibitemShut {NoStop}%
\bibitem [{\citenamefont {Hinshaw}\ \emph {et~al.}(2013)\citenamefont {Hinshaw}
  \emph {et~al.}}]{Hinshaw:2012aka}%
  \BibitemOpen
  \bibfield  {author} {\bibinfo {author} {\bibfnamefont {G.}~\bibnamefont
  {Hinshaw}} \emph {et~al.} (\bibinfo {collaboration} {WMAP}),\ }\href
  {\doibase 10.1088/0067-0049/208/2/19} {\bibfield  {journal} {\bibinfo
  {journal} {Astrophys.J.Suppl.}\ }\textbf {\bibinfo {volume} {208}},\ \bibinfo
  {pages} {19} (\bibinfo {year} {2013})},\ \Eprint
  {http://arxiv.org/abs/1212.5226} {arXiv:1212.5226 [astro-ph.CO]} \BibitemShut
  {NoStop}%
\bibitem [{\citenamefont {Calabrese}\ \emph {et~al.}(2013)\citenamefont
  {Calabrese}, \citenamefont {Hlozek}, \citenamefont {Battaglia}, \citenamefont
  {Battistelli}, \citenamefont {Bond} \emph {et~al.}}]{Calabrese:2013jyk}%
  \BibitemOpen
  \bibfield  {author} {\bibinfo {author} {\bibfnamefont {E.}~\bibnamefont
  {Calabrese}}, \bibinfo {author} {\bibfnamefont {R.~A.}\ \bibnamefont
  {Hlozek}}, \bibinfo {author} {\bibfnamefont {N.}~\bibnamefont {Battaglia}},
  \bibinfo {author} {\bibfnamefont {E.~S.}\ \bibnamefont {Battistelli}},
  \bibinfo {author} {\bibfnamefont {J.~R.}\ \bibnamefont {Bond}},  \emph
  {et~al.},\ }\href {\doibase 10.1103/PhysRevD.87.103012} {\bibfield  {journal}
  {\bibinfo  {journal} {Phys.Rev.}\ }\textbf {\bibinfo {volume} {D87}},\
  \bibinfo {pages} {103012} (\bibinfo {year} {2013})},\ \Eprint
  {http://arxiv.org/abs/1302.1841} {arXiv:1302.1841 [astro-ph.CO]} \BibitemShut
  {NoStop}%
\bibitem [{\citenamefont {Hannestad}\ \emph {et~al.}(2014)\citenamefont
  {Hannestad}, \citenamefont {Hansen},\ and\ \citenamefont
  {Tram}}]{Hannestad:2013ana}%
  \BibitemOpen
  \bibfield  {author} {\bibinfo {author} {\bibfnamefont {S.}~\bibnamefont
  {Hannestad}}, \bibinfo {author} {\bibfnamefont {R.~S.}\ \bibnamefont
  {Hansen}}, \ and\ \bibinfo {author} {\bibfnamefont {T.}~\bibnamefont
  {Tram}},\ }\href {\doibase 10.1103/PhysRevLett.112.031802} {\bibfield
  {journal} {\bibinfo  {journal} {Phys.Rev.Lett.}\ }\textbf {\bibinfo {volume}
  {112}},\ \bibinfo {pages} {031802} (\bibinfo {year} {2014})},\ \Eprint
  {http://arxiv.org/abs/1310.5926} {arXiv:1310.5926 [astro-ph.CO]} \BibitemShut
  {NoStop}%
\bibitem [{\citenamefont {Dasgupta}\ and\ \citenamefont
  {Kopp}(2014)}]{Dasgupta:2013zpn}%
  \BibitemOpen
  \bibfield  {author} {\bibinfo {author} {\bibfnamefont {B.}~\bibnamefont
  {Dasgupta}}\ and\ \bibinfo {author} {\bibfnamefont {J.}~\bibnamefont
  {Kopp}},\ }\href {\doibase 10.1103/PhysRevLett.112.031803} {\bibfield
  {journal} {\bibinfo  {journal} {Phys.Rev.Lett.}\ }\textbf {\bibinfo {volume}
  {112}},\ \bibinfo {pages} {031803} (\bibinfo {year} {2014})},\ \Eprint
  {http://arxiv.org/abs/1310.6337} {arXiv:1310.6337 [hep-ph]} \BibitemShut
  {NoStop}%
\bibitem [{\citenamefont {Bringmann}\ \emph {et~al.}(2014)\citenamefont
  {Bringmann}, \citenamefont {Hasenkamp},\ and\ \citenamefont
  {Kersten}}]{Bringmann:2013vra}%
  \BibitemOpen
  \bibfield  {author} {\bibinfo {author} {\bibfnamefont {T.}~\bibnamefont
  {Bringmann}}, \bibinfo {author} {\bibfnamefont {J.}~\bibnamefont
  {Hasenkamp}}, \ and\ \bibinfo {author} {\bibfnamefont {J.}~\bibnamefont
  {Kersten}},\ }\href {\doibase 10.1088/1475-7516/2014/07/042} {\bibfield
  {journal} {\bibinfo  {journal} {JCAP}\ }\textbf {\bibinfo {volume} {1407}},\
  \bibinfo {pages} {042} (\bibinfo {year} {2014})},\ \Eprint
  {http://arxiv.org/abs/1312.4947} {arXiv:1312.4947 [hep-ph]} \BibitemShut
  {NoStop}%
\bibitem [{\citenamefont {Austermann}\ \emph {et~al.}(2012)\citenamefont
  {Austermann}, \citenamefont {Aird}, \citenamefont {Beall}, \citenamefont
  {Becker}, \citenamefont {Bender} \emph {et~al.}}]{Austermann:2012ga}%
  \BibitemOpen
  \bibfield  {author} {\bibinfo {author} {\bibfnamefont {J.}~\bibnamefont
  {Austermann}}, \bibinfo {author} {\bibfnamefont {K.}~\bibnamefont {Aird}},
  \bibinfo {author} {\bibfnamefont {J.}~\bibnamefont {Beall}}, \bibinfo
  {author} {\bibfnamefont {D.}~\bibnamefont {Becker}}, \bibinfo {author}
  {\bibfnamefont {A.}~\bibnamefont {Bender}},  \emph {et~al.},\ }\href@noop {}
  {\bibfield  {journal} {\bibinfo  {journal} {Proc.SPIE Int.Soc.Opt.Eng.}\
  }\textbf {\bibinfo {volume} {8452}},\ \bibinfo {pages} {84520E} (\bibinfo
  {year} {2012})},\ \Eprint {http://arxiv.org/abs/1210.4970} {arXiv:1210.4970
  [astro-ph.IM]} \BibitemShut {NoStop}%
\bibitem [{\citenamefont {Niemack}\ \emph {et~al.}(2010)\citenamefont
  {Niemack}, \citenamefont {Ade}, \citenamefont {Aguirre}, \citenamefont
  {Barrientos}, \citenamefont {Beall} \emph {et~al.}}]{Niemack:2010wz}%
  \BibitemOpen
  \bibfield  {author} {\bibinfo {author} {\bibfnamefont {M.}~\bibnamefont
  {Niemack}}, \bibinfo {author} {\bibfnamefont {P.}~\bibnamefont {Ade}},
  \bibinfo {author} {\bibfnamefont {J.}~\bibnamefont {Aguirre}}, \bibinfo
  {author} {\bibfnamefont {F.}~\bibnamefont {Barrientos}}, \bibinfo {author}
  {\bibfnamefont {J.}~\bibnamefont {Beall}},  \emph {et~al.},\ }\href@noop {}
  {\bibfield  {journal} {\bibinfo  {journal} {Proc.SPIE Int.Soc.Opt.Eng.}\
  }\textbf {\bibinfo {volume} {7741}},\ \bibinfo {pages} {77411S} (\bibinfo
  {year} {2010})},\ \Eprint {http://arxiv.org/abs/1006.5049} {arXiv:1006.5049
  [astro-ph.IM]} \BibitemShut {NoStop}%
\bibitem [{\citenamefont {Hylen}\ \emph {et~al.}(2003)\citenamefont {Hylen},
  \citenamefont {Bogert}, \citenamefont {Ducar}, \citenamefont {Garkusha} \emph
  {et~al.}}]{Hylen:2002uc}%
  \BibitemOpen
  \bibfield  {author} {\bibinfo {author} {\bibfnamefont {J.}~\bibnamefont
  {Hylen}}, \bibinfo {author} {\bibfnamefont {D.}~\bibnamefont {Bogert}},
  \bibinfo {author} {\bibfnamefont {R.}~\bibnamefont {Ducar}}, \bibinfo
  {author} {\bibfnamefont {V.}~\bibnamefont {Garkusha}},  \emph {et~al.},\
  }\href {\doibase 10.1016/S0168-9002(02)01988-5} {\bibfield  {journal}
  {\bibinfo  {journal} {Nucl.Instrum.Meth.}\ }\textbf {\bibinfo {volume}
  {A498}},\ \bibinfo {pages} {29} (\bibinfo {year} {2003})}\BibitemShut
  {NoStop}%
\end{thebibliography}%

\end{document}